\documentclass[%
 reprint,
 superscriptaddress,
 amsmath,amssymb,
 aps,
]{revtex4-2}

\usepackage{amsmath}
\usepackage{color}
\usepackage{xcolor}
\usepackage{graphicx}
\usepackage{caption}
\usepackage{subcaption}
\usepackage{dcolumn}
\usepackage{bm}
\usepackage[normalem]{ulem}
\usepackage{stackengine}
\newcommand\IncG[2][]{\addstackgap{%
  \raisebox{-.5\height}{\includegraphics[#1]{#2}}}}
\usepackage{enumitem,amssymb}
\usepackage{hyperref}

\makeatletter
\count@=`A \advance\count@\m@ne
\@whilenum\count@<`Z\do{%
  \advance\count@\@ne
  \begingroup\uccode`a=\count@
  \uppercase{\endgroup\DeclareMathSymbol{a}}{\mathalpha}{operators}{\count@}%
}
\makeatother

\makeatletter
\count@=`a \advance\count@\m@ne
\@whilenum\count@<`z\do{%
  \advance\count@\@ne
  \begingroup\uccode`a=\count@
  \uppercase{\endgroup\DeclareMathSymbol{a}}{\mathalpha}{operators}{\count@}%
}
\makeatother

\newcommand{\revf}[1]{{\color{black}{#1}}}

\newcommand{\mth}{M$_{\rm{th}}$}
\newcommand{\mmax}{M$_{\rm{max}}$}

\newcommand{\cmax}{C$_{\rm{max}}$}

\begin{document}

\preprint{APS/123-QED}

\title{Numerical relativity simulations of prompt collapse mergers: threshold mass and phenomenological constraints on neutron star properties after GW170817}

\author{Rahul Kashyap}
\email{rkk5314@psu.edu}
\affiliation{Institute for Gravitation and the Cosmos, The Pennsylvania State University, University Park, PA 16802, USA}
\affiliation{Department of Physics, The Pennsylvania State University, University Park, PA 16802, USA}

\author{Abhishek Das}%
\affiliation{Institute for Gravitation and the Cosmos, The Pennsylvania State University, University Park, PA 16802, USA}
\affiliation{Department of Physics, The Pennsylvania State University, University Park, PA 16802, USA}

\author{David Radice} 
\affiliation{Institute for Gravitation and the Cosmos, The Pennsylvania State University, University Park, PA 16802, USA}
\affiliation{Department of Physics, The Pennsylvania State University, University Park, PA 16802, USA}
\affiliation{Department of Astronomy \& Astrophysics, The Pennsyvlania State University, University Park, PA 16802, USA}

\author{Surendra Padamata}
\affiliation{Institute for Gravitation and the Cosmos, The Pennsylvania State University, University Park, PA 16802, USA}
\affiliation{Department of Physics, The Pennsylvania State University, University Park, PA 16802, USA}

\author{Aviral Prakash}
\affiliation{Institute for Gravitation and the Cosmos, The Pennsylvania State University, University Park, PA 16802, USA}
\affiliation{Department of Physics, The Pennsylvania State University, University Park, PA 16802, USA}

\author{Domenico Logoteta}
\affiliation{Dipartimento di Fisica, Universit\`{a} di Pisa, Largo B.  Pontecorvo, 3 I-56127 Pisa, Italy}
\affiliation{INFN, Sezione di Pisa, Largo B. Pontecorvo, 3 I-56127 Pisa, Italy}

\author{Albino Perego}
\affiliation{Dipartimento di Fisica, Università di Trento, Via Sommarive 14, 38123 Trento, Italy}
\affiliation{INFN-TIFPA, Trento Institute for Fundamental Physics and Applications, ViaSommarive 14, I-38123 Trento, Italy}

\author{Daniel A. Godzieba}
\affiliation{Department of Physics, The Pennsylvania State University, University Park, PA 16802, USA}

\author{Sebastiano Bernuzzi}
\affiliation{Theoretisch-Physikalisches Institut, Friedrich-Schiller Universit\"{a}t Jena, 07743, Jena, Germany}

\author{Ignazio Bombaci}
\affiliation{Dipartimento di Fisica, Universit\`{a} di Pisa, Largo B.  Pontecorvo, 3 I-56127 Pisa, Italy}
\affiliation{INFN, Sezione di Pisa, Largo B. Pontecorvo, 3 I-56127 Pisa, Italy} 

\author{Farrukh J. Fattoyev}
\affiliation{Department of Physics, Manhattan College, Riverdale, NY 10471, USA}

\author{Brendan T. Reed}
\affiliation{Department of Astronomy, Indiana University, Bloomington, IN 47405, USA}

\author{Andr\'e da Silva Schneider}
\affiliation{The Oskar Klein Centre, Department of Astronomy,\\ Stockholm University, AlbaNova, SE-106 91 Stockholm, Sweden}

\date{\today}


\begin{abstract}
We determine the threshold mass for prompt (no bounce) black hole formation in equal-mass neutron star (NS) mergers using a new set of 227 numerical relativity simulations. We consider 23 phenomenological and microphysical finite temperature equations of state (EOS), including models with hyperons and first-order phase transitions to deconfined quarks. We confirm the existence of EOS-insensitive relations between the threshold mass, the binary tidal parameter at the threshold ($\Lambda_{th}$), the maximum mass of nonrotating NSs, and the radii of reference mass NSs. 
We combine the EOS-insensitive relations, phenomenological constraints on NS properties and observational data from GW170817 to derive an improved lower limit on radii of maximum mass and 1.6 M$_\odot$ NS of 9.81~km and 10.90~km, respectively. We also constrain the radius and quadrupolar tidal deformability ($\Lambda$) of a 1.4 $M_\odot$ NS to be larger than 10.74~km and 172, respectively. We consider uncertainties in all independent parameters -- fitting coefficients as well as GW170817 masses while reporting the range of radii constraints. We \revf{discuss an approach} to constrain the upper as well as lower limit of NS maximum mass using future BNS detections and their identification as prompt or delayed collapse. With future observations it will be possible to derive even tighter constraints on the properties of matter at and above nuclear density using the method proposed in this work.
\end{abstract}

\maketitle

\section{Introduction}
\label{sec:intro}

Binary neutron star (BNS) mergers are one of the most important events in gravitational wave astronomy. At least two such events have been detected by LIGO and Virgo thus far (namely GW170817~\cite{TheLIGOScientific:2017qsa} and GW190425~\cite{Abbott:2020uma}). Apart from gravitational waves, BNS mergers may produce electromagnetic (EM) counterparts across the entire EM spectrum which can be detected by various space-based and ground-based observatories. These events result in one of the following two outcomes. The merger remnant may be a massive, differentially rotating neutron star secured by centrifugal and thermal effects that possibly collapses to a black hole on a dynamical or secular timescale. Material that becomes gravitationally unbound during the coalescence undergoes rapid neutron-capture nucleosynthesis and contribute to the galactic enrichment by heavy elements. The energy released by the radioactive decay of the nucleosynthesis products power electromagnetic counterparts, called kilonova \citep{Li:1998bw, Kulkarni:2005jw, Metzger:2010sy, Kasen:2013xka, Tanaka:2013ana, Metzger:2019zeh, Hotokezaka:2021ofe}. Alternatively, the remnant may form a black hole immediately upon merger, the so-called prompt collapse. In this case, if the stars have comparable masses, most of the matter falls immediately into the black hole resulting in an EM-quiet merger, so the occurrence of prompt collapse can be determined with multi-messenger observations \revf{ \cite{Shibata:2005ss, Hotokezaka:2011dh, Hotokezaka:2012ze, Bauswein:2013yna, Radice:2017lry, Radice:2018pdn, Kiuchi:2019lls}.} Bright electromagnetic counterparts might still be expected for high mass-ratio prompt-collapse binaries \cite{Bernuzzi:2020txg}. This study is concerned with the binary threshold mass \mth{} that separates these two outcomes for equal mass binaries.

The phenomenon of prompt collapse has been investigated by several groups~\cite{Shibata:2005ss, Hotokezaka:2011dh, Bauswein:2013jpa, Zappa:2017xba, Koppel:2019pys, Bauswein:2020aag, Bauswein:2020xlt}. It is widely accepted that the threshold mass for prompt collapse should be strongly correlated with other physical properties of the equation of state (EOS). This can be used to place constraints on the EOS using information from future possible observations of prompt and/or delayed collapse. \citet{Shibata:2005ss} first proposed that the minimum total mass of binary undergoing prompt collapse is directly proportional to the maximum mass \mmax{} of cold non-rotating neutron stars. This was later corroborated with an extensive study spanning multiple EOSs by \citet{Hotokezaka:2011dh}. \citet{Bauswein:2013jpa} extended the theoretical correlation study by proposing a further linear relationship between the compactness of the maximum mass neutron star (\cmax{}=$GM_{max}/R_{max}c^2$ where G is the Newton's gravitational constant, c is the speed of light, $M_{max}$ and $R_{max}$ are the mass and radius of the maximum mass NS) and the proportionality constant ($\rm{k}_{\rm{th}} = \rm{M}_{\mathrm{th}}/\rm{M}_{\mathrm{max}}$) between the threshold mass (\mth) and maximum mass (\mmax). They also found new EOS-insensitive relations between $k_{th}$ and other quantities depending on the EOS, such as with the modified compactness, C$^*_{1.6}$($=G \rm{M}_{\max}/c^2 R_{1.6}$ where $R_{1.6}$ is the radius of a 1.6 $M_\odot$ NS) \cite{Bauswein:2013jpa}, radii and quadrupolar tidal polarizability parameter \cite{Damour:2012yf} (hereafter, shortly, tidal deformability) ($\Lambda$) at few particular values of masses \cite{Bauswein:2020xlt}. Since GW170817 had a bright EM counterpart \cite{LIGOScientific:2017ync} it is widely believed not to have been a prompt collapse event, e.g., \cite{Margalit:2017dij}. Using this information and the proposed correlations between $k_{th}$ and $C_{1.6}^*$, \citet{Bauswein:2017vtn} have derived a lower limit for $R_{1.6}$, the radius of a cold $1.6\ M_\odot$ NS. \citet{Koppel:2019pys} have used a nonlinear fit between \mmax{} and \mth{} motivated by the condition that $\rm{k}_{\rm{th}} \rightarrow 0$ as compactness reaches that of BH. They derive a lower limit of radii as a function of NS masses later extended by \citet{Tootle:2021umi} to asymmetric binaries. \citet{Agathos:2019sah} presented a Bayesian framework based on these correlations to calculate the probability of prompt collapse for a given merger from the inspiral GW signal. More recently, \citet{Bauswein:2020aag} and \citet{Bauswein:2020xlt} have considered the effect of mass-ratio and phase transitions. They claimed that the combined measurement of \mth{} and the binary tidal parameter of the corresponding binary, $\tilde\Lambda_{th}$, could reveal the presence of QCD phase transitions in cold, dense matter. In particular, they identified a region in the $M_{th}-\tilde\Lambda_{th}$ plane that was only populated by EOS models with strong first-order phase transitions. They argued that, should the observationally determined \mth{} and $\tilde\Lambda_{th}$ lay in this region, this would be smoking gun evidence for a phase transition. \\

\revf{Recently, several studies have investigated the impact of mass ratio on the \mth{} \citep{Bauswein:2020xlt,Perego:2021mkd,Kolsch:2021lub,Tootle:2021umi}. \citet{Bauswein:2020xlt} have found that the \mth{} may decrease or increase for asymmetric systems depending upon the stiffness of EOSs.}
In \citet{Bauswein:2020xlt}, a fitting formula of the difference is provided with respect to the non-rotating NS properties along with an explanation of this difference using the angular momentum of binaries. Using the fitting procedure of \citet{Koppel:2019pys}, \citet{Tootle:2021umi} have extended their earlier studies by looking at the impacts of mass ratio and spin. \citet{Perego:2021mkd} have looked at the effect of mass ratio and provided an explanation from the fundamental perspective of nuclear physics and angular momentum of binaries. They provide a broken linear fit of \mth{} with respect to mass ratio in two regimes (lower and higher than $q=0.725$) and incompressibility at the maximum NS density ($K_{max}$), fundamental to the behavior of EOS. They further provide a method to constrain the $K_{max}$ by observing the difference of \mth{} between symmetric and non-symmetric 
 BNS systems. \citet{Kolsch:2021lub} discusses the impact of asymmetry by introducing an extended version of the fitting formula from \citet{Bauswein:2020xlt}
 to account for the different behaviors in two regimes of mass ratio and for different EOS. There is an agreement among these studies that, while spin can 
 increase the threshold mass by 5-10\%, the effect of large asymmetry is to decrease the threshold mass by up to 8\% for most of the EOSs. However, there are few EOSs for which \mth{} increases for intermediate mass ratios \citep{Perego:2021mkd}.

In this work, we revisit the issue of the prompt BH formation in binary NS mergers. On one hand, we confirm the existence of EOS-insensitive relations between $k_{th}$ and other parameters that depend on the EOS. On the other hand, we find small, but statistically significant, systematic deviations between our results, obtained with full general-relativity simulations, and those of \revf{ \citet{Bauswein:2013jpa} that used approximate general relativity simulations. However, we find our linear fit to be in closer agreement with \citet{Bauswein:2020xlt} with differences for data points at higher values of $C_{max}$.} Our results are consistent with those obtained by other groups also performing general-relativity simulations. We \revf{extend the methods of \citet{Bauswein:2017vtn} to} combine observational data and simulation results and derive updated constraints on the radius of $1.4\ M_\odot$ and $1.6\ M_\odot$ NSs, as well as on the radius of the maximum mass NS. We also re-examine the claim of \citet{Bauswein:2020aag} and \citet{Bauswein:2020xlt} that particular combination of values of \mth and $\Lambda_{th}$ would imply the presence of a first-order phase transition in the core of NSs. We find consistency with their claim for most of the EOS, however the error bars on $\Lambda_{th}$ (the tidal deformability of a NS with mass equal to half of $M_{th}$) of few nucleonic EOS shows large uncertainties requiring future high resolution studies to confirm or refute their hypothesis.

In the next section, we describe the EOS models used in our simulation classified into different categories. We describe the details of numerical simulations and the method to identify prompt collapse in section \ref{sec:num_rel}. Further, we describe three main classes of outcomes from BNS mergers using the midplane rest mass density slices and time series of key physical quantities in section \ref{sec:results}. In the same section, we describe the correlation between \mth{} and \mmax{}, the phenomenological constraints on  NS properties and derive constraints on \mmax{}, radii and \mth{} using the correlations. Finally, we provide a brief conclusion of our results in section \ref{sec:conclusions}. 

\section{Equation of state}
\label{sec:EOS_models}

In the present work we consider a set of 23 EOS models. This set includes: 
(\rm{a}) 21 composition-dependent EOSs derived using different frameworks to describe the many-body dynamics of the stellar constituents; 
(\rm{b}) two entirely phenomenological zero-temperature piecewise polytropic EOS from \cite{Godzieba:2020bbz}. 

Fifteen EOSs of the subset (\rm{a}) are nucleonic models, i.e. they include only neutrons and protons as hadronic components of dense matter.  More specifically the nucleonic EOSs we consider are: 
BLh \cite{Bombaci:2018ksa,Logoteta:2020yxf},  
HS(DD2) \cite{Typel:2009sy, Hempel:2009mc}, LS220 \cite{Lattimer:1991nc}, SFHo \cite{Steiner:2012rk}, 
SRO(SLy) \cite{Douchin:2001sv, Schneider:2017tfi}, 9 variants of the SRO EOS with different values of the empirical nuclear parameters \cite{Schneider:2017tfi, Schneider:2019shi} and 
the Big Apple EOS (BA from now on) \cite{Fattoyev:2020cws}. With the exception of the BA EOS, all these models are finite-temperature EOSs, i.e. thermal effects are consistently calculated within the adopted many-body framework. 
Three EOS models include hyperons in addition to nucleons: HS(BHB$\Lambda\phi$, just BHB elsewhere in this paper) \cite{Banik:2014qja}, 
H3 and H4 \cite{Glendenning:1991es, Lackey:2005tk, Read:2008iy}. The latter two are zero-temperature models. 
Finally we consider three EOS models that include a transition to a phase with deconfined quark: the finite-temperture  BLQ \cite{Prakash:2021wpz} and  DD2qG (Logoteta et al., in prep) EOSs, and the 
zero-temperature ALF2 \cite{Alford:2004pf} EOS. 
In the BLQ EOS the nucleonic phase is described by the BLh model \cite{Logoteta:2020yxf,Bombaci:2018ksa} 
while the quark phase is described by an extended version of the phenomenological bag model EOS which includes the effects of gluon mediated QCD interactions between quarks up to the second order in the QCD coupling $\alpha_s$ \cite{2001PhRvD..63l1702F, 2005ApJ...629..969A, 2011ApJ...740L..14W}. 
The nucleonic and the quark phase are then joint assuming a first order phase transition according to the so called Gibbs construction \cite{Glendenning:1992vb}. 
The possibility of a quark deconfinement phase transition is again considered in the DD2qG and ALF2 EOS models and the Gibbs construction is still adopted in both cases to join the hadronic and the quark phases. For the latter phase, in the case of the DD2qG, the same quark model used for the BLQ EOS is employed, while in ALF2 the quark model was extended to include the effect of  color-superconducting assuming a color-flavor-locked quark phase \cite{Alford:2004pf}. 
 
Finally we include thermal effects in the zero temperature EOS models (BA, H3, H4, ALF2, and 
piecewise polytropic EOSs from subset (\rm{b}), GRW1 and GRW2) by adding a thermal contribution with adiabatic index 
$\Gamma_{th} = 1.7$ following, e.g., Refs.~\cite{Shibata:2005ss, Bauswein:2010dn, Endrizzi:2018uwl, Figura:2020fkj}, 
see \cite{Raithel:2019gws, Raithel:2021hye} for an alternative approach. 

\begin{figure*} 
    \includegraphics[width=\linewidth]{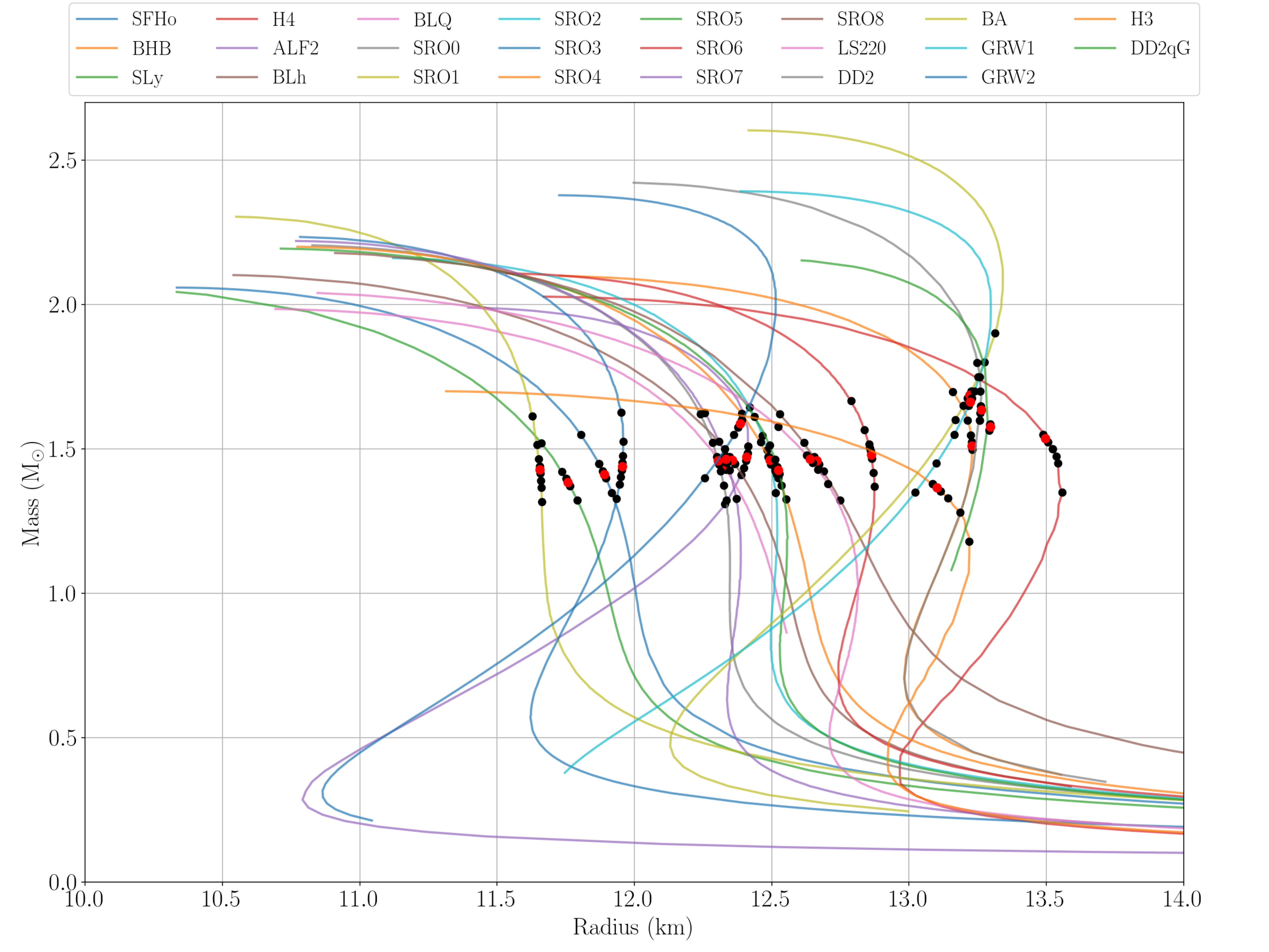}
    \caption{Mass-radius curves for the equations of state used in our simulations. The dots indicate component masses and radii for binaries that we simulated, the red ones denoting the minimum mass binaries that underwent prompt collapse.}
    \label{fig:EOS}
\end{figure*}

The mass radius curves for non-rotating neutron stars obtained for all the considered EOS models are shown 
in Fig.~\ref{fig:EOS}.

\section{Numerical Simulations}
\label{sec:num_rel}

We construct irrotational quasi-circular binary initial data with the pseudospectral code \texttt{Lorene} \cite{Gourgoulhon:2000nn}. The initial separation between the centers of the two stars is typically taken to be $40\ {\rm km}$, the exception being the BLh initial data, used for both the BLh and BLQ binaries, for which the initial separation is $45\ {\rm km}$ \cite{Prakash:2021wpz}. Evolutions are carried out with the \texttt{WhiskyTHC} code \cite{Radice:2012cu, Radice:2013hxh, Radice:2013xpa, Radice:2015nva, Radice:2016dwd, Radice:2017zta, Radice:2018pdn}, which is built on top of the \texttt{Einstein Toolkit} \cite{Loffler:2011ay}. Our simulations make use of the \texttt{Carpet} adaptive mesh refinement (AMR) framework \cite{Schnetter:2003rb, Reisswig:2012nc}, which implements the Berger-Oliger scheme with refluxing \cite{Berger:1984zza, 1989JCoPh..82...64B}. We use 7 levels of mesh refinement, with the finest grid covering the NSs and the merger remnant entirely. Our fiducial simulations, referred to as the standard resolution (SR), have a resolution of $0.125 G M_\odot / c^2 {\simeq}185\ {\rm m}$. Additional simulations are performed at the lower resolution (LR) of $0.167  G M_\odot / c^2 {\simeq} 246\ {\rm m}$ for all the binaries near the threshold mass. The grid setup is discussed in detail in Ref.~\cite{Radice:2018pdn}. Simulations performed with microphysical EOS also account for neutrino emission using the leakage scheme discussed in \cite{Galeazzi:2013mia, Radice:2016dwd}. 

We define a binary to have undergone prompt collapse if the remnant does not bounce back after merger and instead a BH is immediately formed. We remark that this definition has been used in most previous studies, e.g., \cite{Hotokezaka:2011dh, Bauswein:2013jpa, Kiuchi:2019lls}, with the notable exceptions of \citet{Koppel:2019pys} and \citet{Tootle:2021umi}, which instead use a condition based on the time for the formation of a BH after merger. As in most previous studies, but again differently from \citet{Koppel:2019pys}, the BH formation threshold is obtained using a bracketing procedure. \citet{Koppel:2019pys} used an extrapolation method to identify \mth. For most EOSs, we are able to determine \mth{} to within $0.05\ M_\odot$. For those EOSs for which we find discrepancies between standard and low resolution simulations (BA, DD2, SRO5 and SFHo), we have reported the threshold masses corresponding to both resolutions in table \ref{tab:main}. In our analysis, we have used SR results, but we have extended the error bars to account for this discrepancy. Additionally, we have checked that our results are robust, to within the estimated errors, with respect to changes in the gauge conditions for the BLQ equation of state where the difference between Gamma and integrated Gamma driver (GD and IGD, respectively) gauge conditions \citep{Reisswig:2013sqa,Pollney:2009yz} is equal to the error bar in \mth{} and is non-existent for BLh. Overall, we have performed 227 numerical relativity simulations for this study.

\begin{table*} 
\begin{center} 
\caption{\label{tab:main} Physical properties of each EOS used in this study along with properties of NS near threshold. Res. is the resolutions used for each EOS.} 
\label{table: main} 
\scalebox{1.02}{ 
 \setlength{\tabcolsep}{5pt} 
 \begin{tabular}{ c c c c  c  c  c  c  c  c  c } 
 \hline\hline 
 \\[-0.9em] 
 EOS & Res. & Gauge & k$_{\rm{th}}$ & C$_{\rm{max}}$ & M$_{\rm{g,max}}$ & C$_{\rm{th}}$ & $\Lambda_{\rm{th}}$ & $\Lambda_{1.4}$ & R$_{1.4}$ & R$_{1.6}$ \\ 
  & & & & & $(M_\odot)$ & & & & (km) & (km) \\
 \hline 
 \hline 
\rule{0pt}{4ex}SFHo & SR & IGD & 1.37$~\pm~0.01$ & 0.29 & 2.06 & 0.18$~\pm~0.01$ & 312.94$~\pm~68.34$ & 332.67 & 11.90 & 11.77 \\ 
 & LR &  & 1.39$~\pm~0.01$ & & & & & & & \\
\rule{0pt}{4ex}BHB & SR, LR & IGD &  1.44$~\pm~0.01$ & 0.27 & 2.10 & 0.17$~\pm~0.00$ & 466.59$~\pm~50.89$ & 754.43 & 13.22 & 13.21 \\ 
\rule{0pt}{4ex}SLy & SR, LR & IGD &  1.36$~\pm~0.01$ & 0.30 & 2.05 & 0.18$~\pm~0.00$ & 314.40$~\pm~39.21$ & 310.44 & 11.75 & 11.59 \\ 
\rule{0pt}{4ex}H4 & SR, LR & IGD &  1.52$~\pm~0.01$ & 0.26 & 2.03 & 0.17$~\pm~0.00$ & 487.06$~\pm~59.06$ & 933.11 & 13.55 & 13.45 \\ 
\rule{0pt}{4ex}ALF2 & SR, LR & IGD &  1.48$~\pm~0.01$ & 0.26 & 1.99 & 0.18$~\pm~0.00$ & 436.98$~\pm~51.19$ & 607.18 & 12.39 & 12.41 \\ 
\rule{0pt}{4ex}BLh & SR, LR & IGD &  1.39$~\pm~0.01$ & 0.30 & 2.10 & 0.17$~\pm~0.00$ & 321.97$~\pm~36.59$ & 430.29 & 12.42 & 12.24 \\ 
 &  & GD & 1.39$~\pm~0.01$ & & & & & & & \\
\rule{0pt}{4ex}BLQ & SR, LR & IGD &  1.44$~\pm~0.01$ & 0.28 & 1.99 & 0.17$~\pm~0.00$ & 368.83$~\pm~40.64$ & 434.57 & 12.40 & 12.23 \\ 
 &  & GD & 1.43$~\pm~0.01$ & & & & & & & \\
\rule{0pt}{4ex}SRO0 & SR, LR & IGD &  1.34$~\pm~0.01$ & 0.30 & 2.21 & 0.18$~\pm~0.00$ & 317.18$~\pm~33.55$ & 433.19 & 12.33 & 12.26 \\ 
\rule{0pt}{4ex}SRO1 & SR, LR & IGD &  1.25$~\pm~0.01$ & 0.32 & 2.30 & 0.18$~\pm~0.00$ & 258.95$~\pm~27.60$ & 309.37 & 11.67 & 11.64 \\ 
\rule{0pt}{4ex}SRO2 & SR, LR & IGD &  1.36$~\pm~0.01$ & 0.29 & 2.16 & 0.17$~\pm~0.00$ & 346.86$~\pm~36.99$ & 474.57 & 12.51 & 12.45 \\ 
\rule{0pt}{4ex}SRO3 & SR, LR & IGD &  1.30$~\pm~0.01$ & 0.31 & 2.23 & 0.18$~\pm~0.00$ & 313.79$~\pm~32.87$ & 393.73 & 11.96 & 11.96 \\ 
\rule{0pt}{4ex}SRO4 & SR, LR & IGD &  1.34$~\pm~0.01$ & 0.30 & 2.20 & 0.17$~\pm~0.00$ & 341.54$~\pm~37.01$ & 478.06 & 12.53 & 12.41 \\ 
\rule{0pt}{4ex}SRO5 & SR & IGD &  1.31$~\pm~0.01$ & 0.30 & 2.19 & 0.17$~\pm~0.01$ & 400.08$~\pm~81.77$ & 474.34 & 12.54 & 12.45 \\ 
 & LR &  & 1.33$~\pm~0.01$ & & & & & & & \\
\rule{0pt}{4ex}SRO6 & SR & IGD &  1.41$~\pm~0.01$ & 0.27 & 2.11 & 0.17$~\pm~0.00$ & 392.15$~\pm~62.18$ & 589.98 & 12.87 & 12.82 \\ 
 & LR &  & 1.44$~\pm~0.01$ & & & & & & & \\
\rule{0pt}{4ex}SRO7 & SR, LR & IGD &  1.33$~\pm~0.01$ & 0.31 & 2.22 & 0.18$~\pm~0.00$ & 327.70$~\pm~35.91$ & 463.74 & 12.36 & 12.28 \\ 
\rule{0pt}{4ex}SRO8 & SR, LR & IGD &  1.35$~\pm~0.01$ & 0.30 & 2.18 & 0.17$~\pm~0.00$ & 340.46$~\pm~38.82$ & 473.99 & 12.71 & 12.55 \\ 
\rule{0pt}{4ex}LS220 & SR, LR & IGD &  1.45$~\pm~0.01$ & 0.28 & 2.04 & 0.17$~\pm~0.00$ & 377.03$~\pm~44.93$ & 547.43 & 12.69 & 12.50 \\ 
\rule{0pt}{4ex}DD2 & SR & IGD &  1.35$~\pm~0.01$ & 0.30 & 2.42 & 0.18$~\pm~0.01$ & 300.42$~\pm~62.34$ & 769.37 & 13.23 & 13.27 \\ 
 & LR &  & 1.33$~\pm~0.01$ & & & & & & & \\
\rule{0pt}{4ex}BA & SR & IGD &  1.30$~\pm~0.03$ & 0.31 & 2.60 & 0.19$~\pm~0.01$ & 265.15$~\pm~63.33$ & 739.03 & 13.04 & 13.18 \\ 
 & LR &  & 1.32$~\pm~0.01$ & & & & & & & \\
\rule{0pt}{4ex}GRW1 & SR, LR & IGD &  1.39$~\pm~0.01$ & 0.29 & 2.39 & 0.19$~\pm~0.00$ & 306.81$~\pm~28.41$ & 818.98 & 13.07 & 13.20 \\ 
\rule{0pt}{4ex}GRW2 & SR, LR & IGD &  1.34$~\pm~0.01$ & 0.30 & 2.38 & 0.19$~\pm~0.00$ & 271.05$~\pm~25.40$ & 553.33 & 12.26 & 12.40 \\ 
\rule{0pt}{4ex}H3 & SR, LR & IGD &  1.61$~\pm~0.01$ & 0.22 & 1.70 & 0.15$~\pm~0.00$ & 797.22$~\pm~117.74$ & 655.56 & 13.06 & 12.48 \\ 
\rule{0pt}{4ex}DD2qG & SR & IGD &  1.47$~\pm~0.01$ & 0.25 & 2.15 & 0.18$~\pm~0.00$ & 354.52$~\pm~30.51$ & 690.24 & 13.27 & 13.29 \\ 
\hline 
\end{tabular} 
}
\end{center} 
\end{table*}

\section{Results}
\label{sec:results}

\begin{figure*}[!t]
\begin{tabular}{p{0\textwidth}p{.3\textwidth}p{.3\textwidth}p{.3\textwidth}p{.1\textwidth}}
&\IncG[trim={1cm 0 1cm 0}, width=.35\textwidth,height=.3\textwidth]{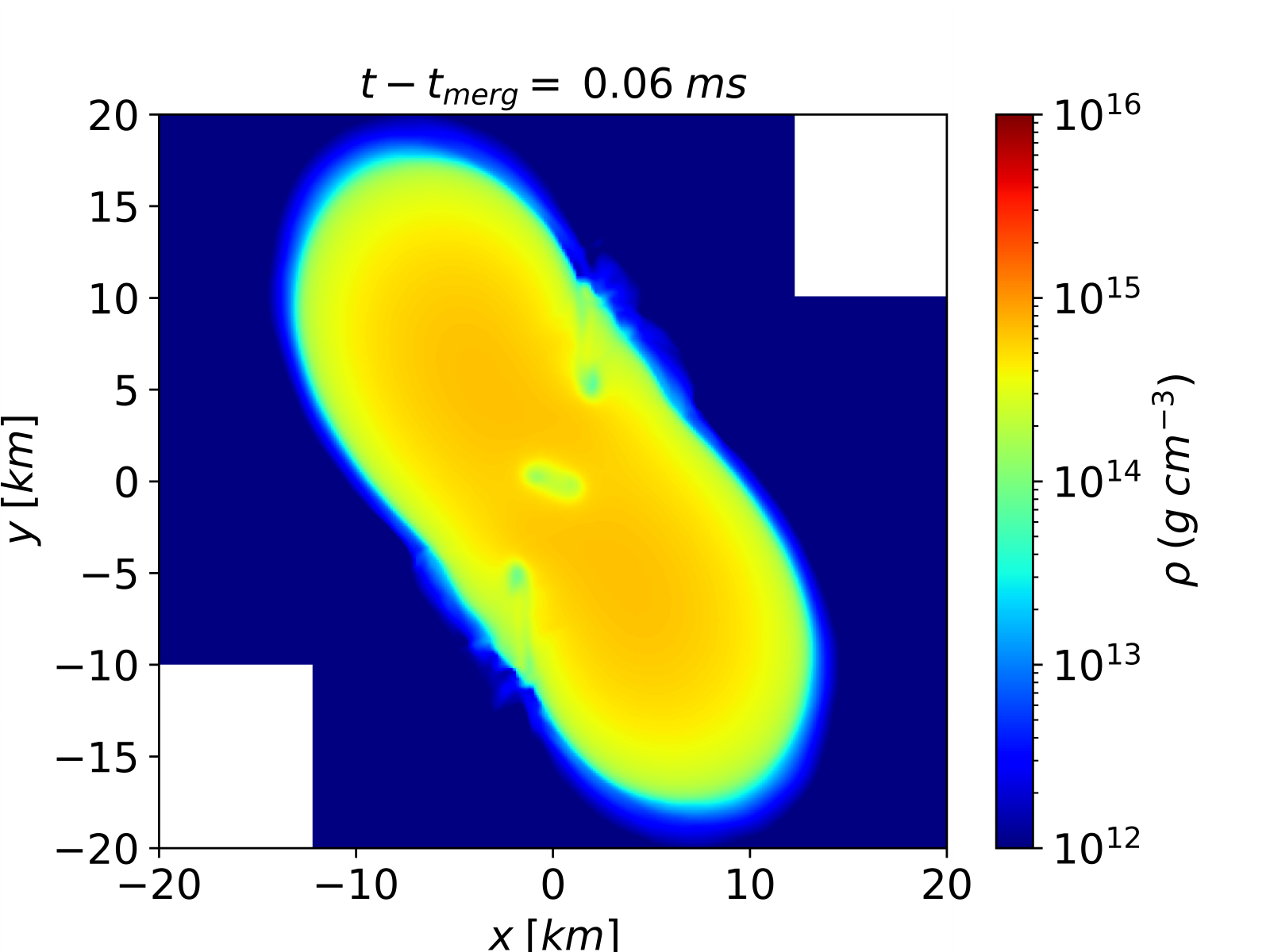}
&\IncG[trim={1cm 0 1cm 0}, width=.35\textwidth,height=.3\textwidth]{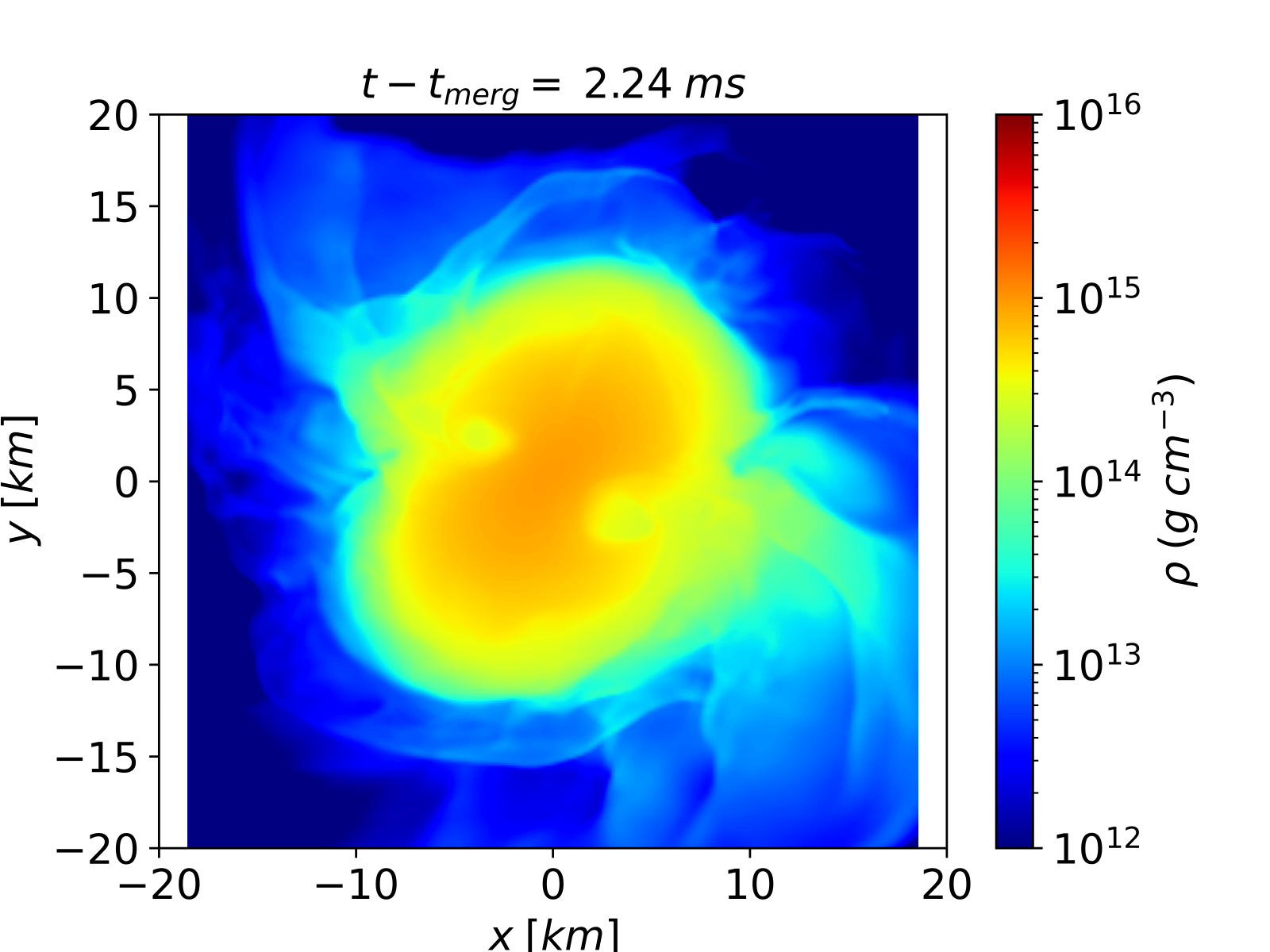}
&\IncG[trim={1cm 0 0 0}, width=.35\textwidth,height=.3\textwidth]{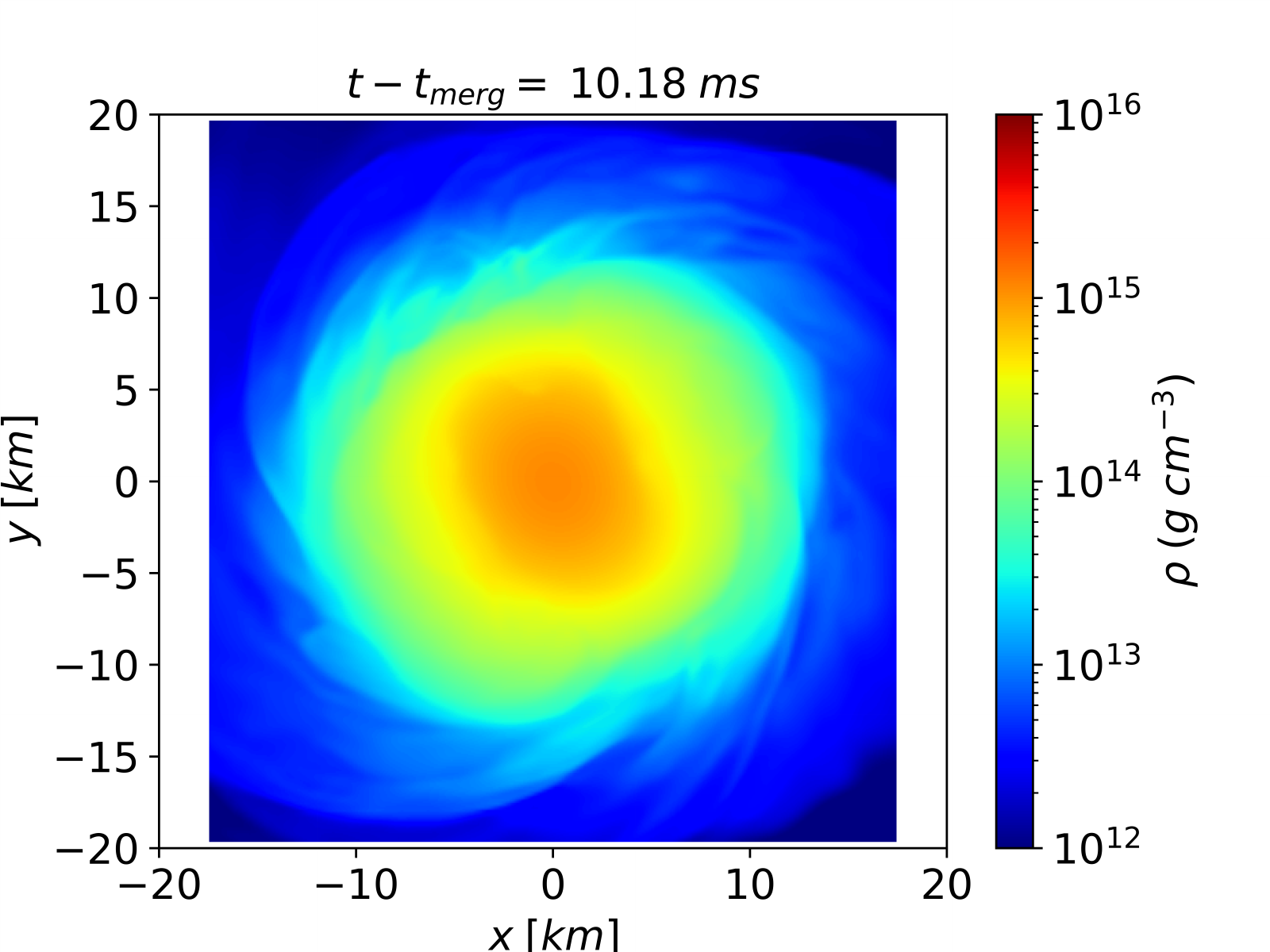}\\
&\IncG[trim={1cm 0 1cm 0}, width=.35\textwidth,height=.3\textwidth]{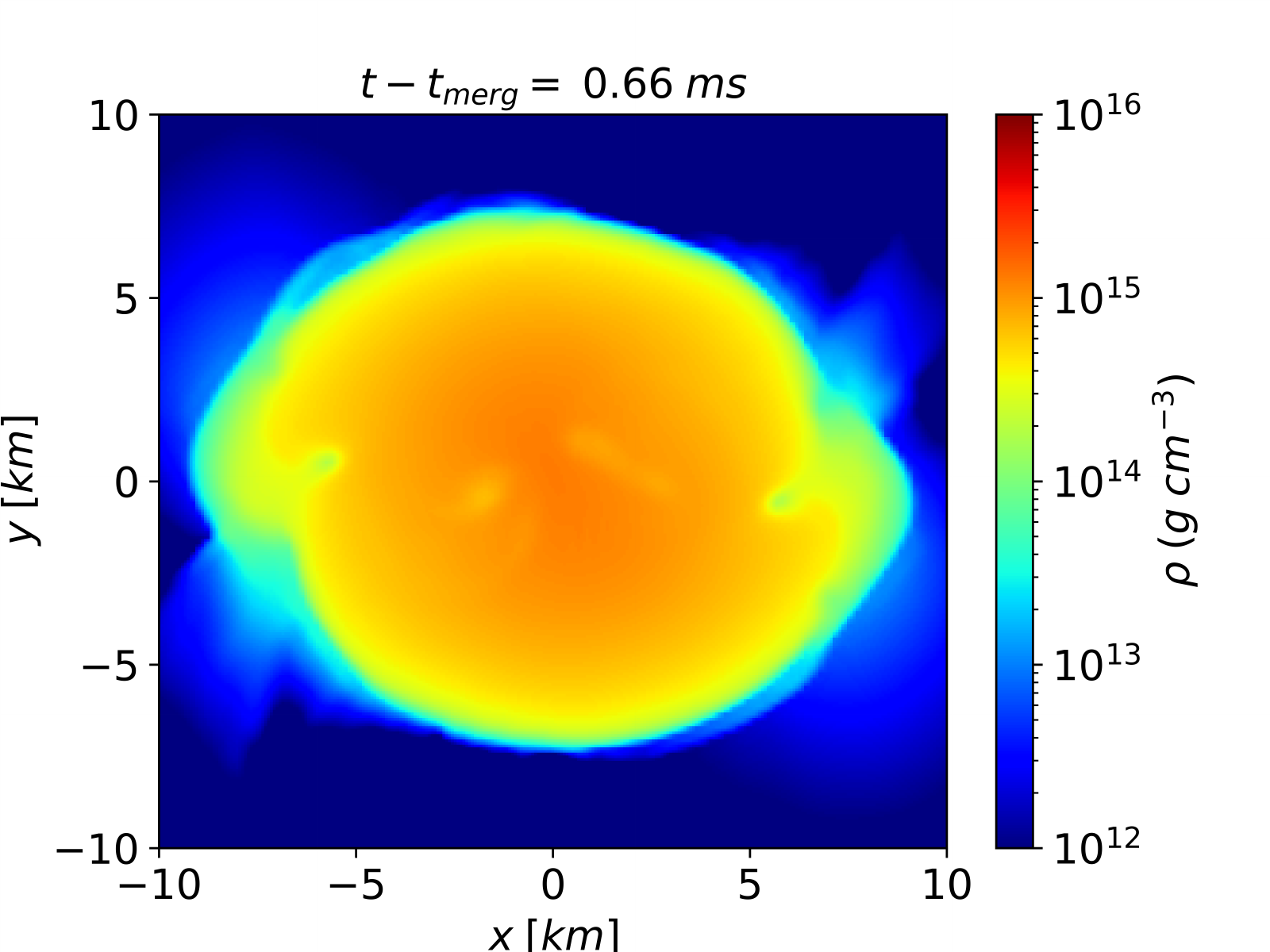}
&\IncG[trim={1cm 0 1cm 0}, width=.35\textwidth,height=.3\textwidth]{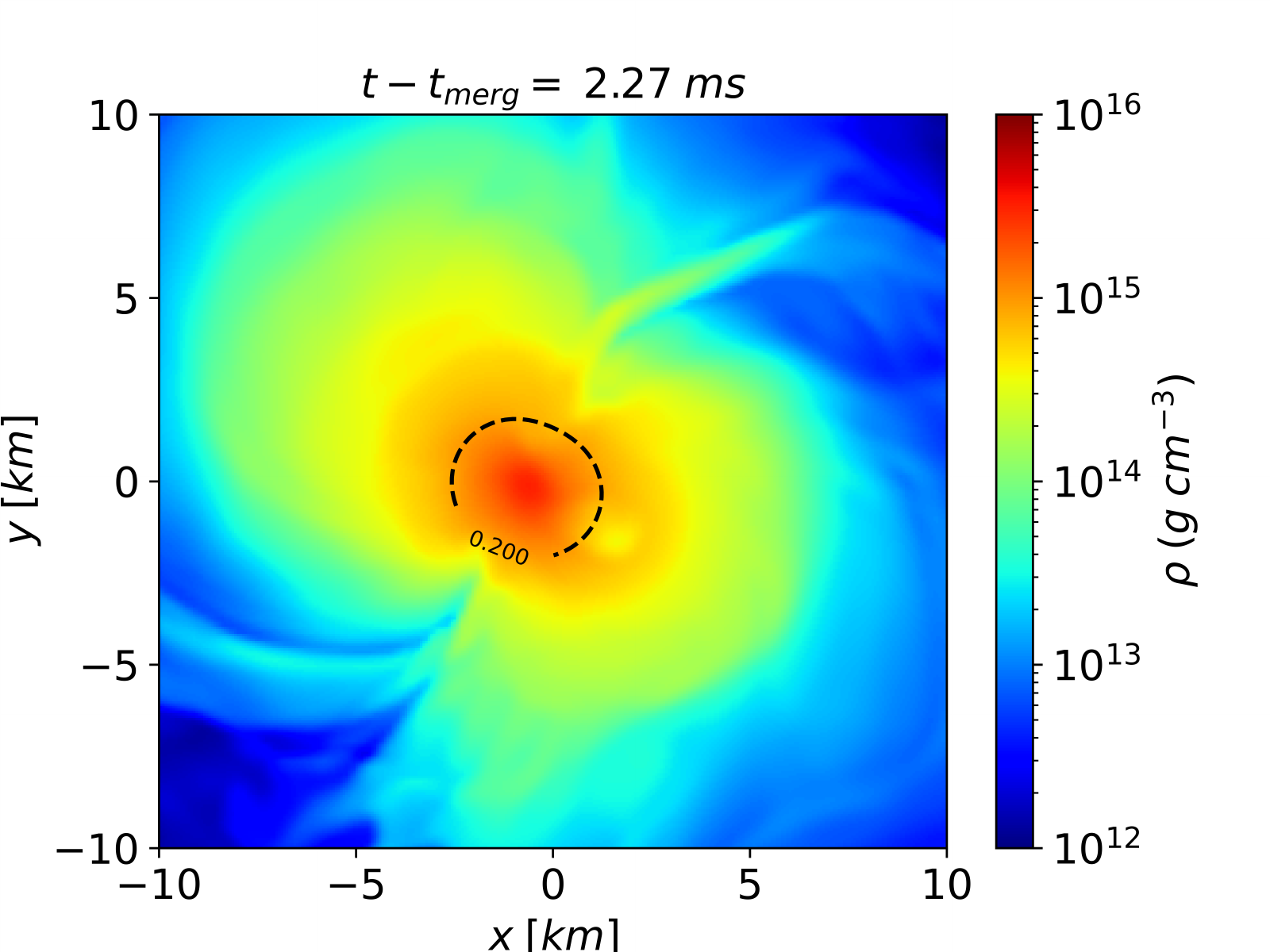}
&\IncG[trim={1cm 0 0 0}, width=.35\textwidth,height=.3\textwidth]{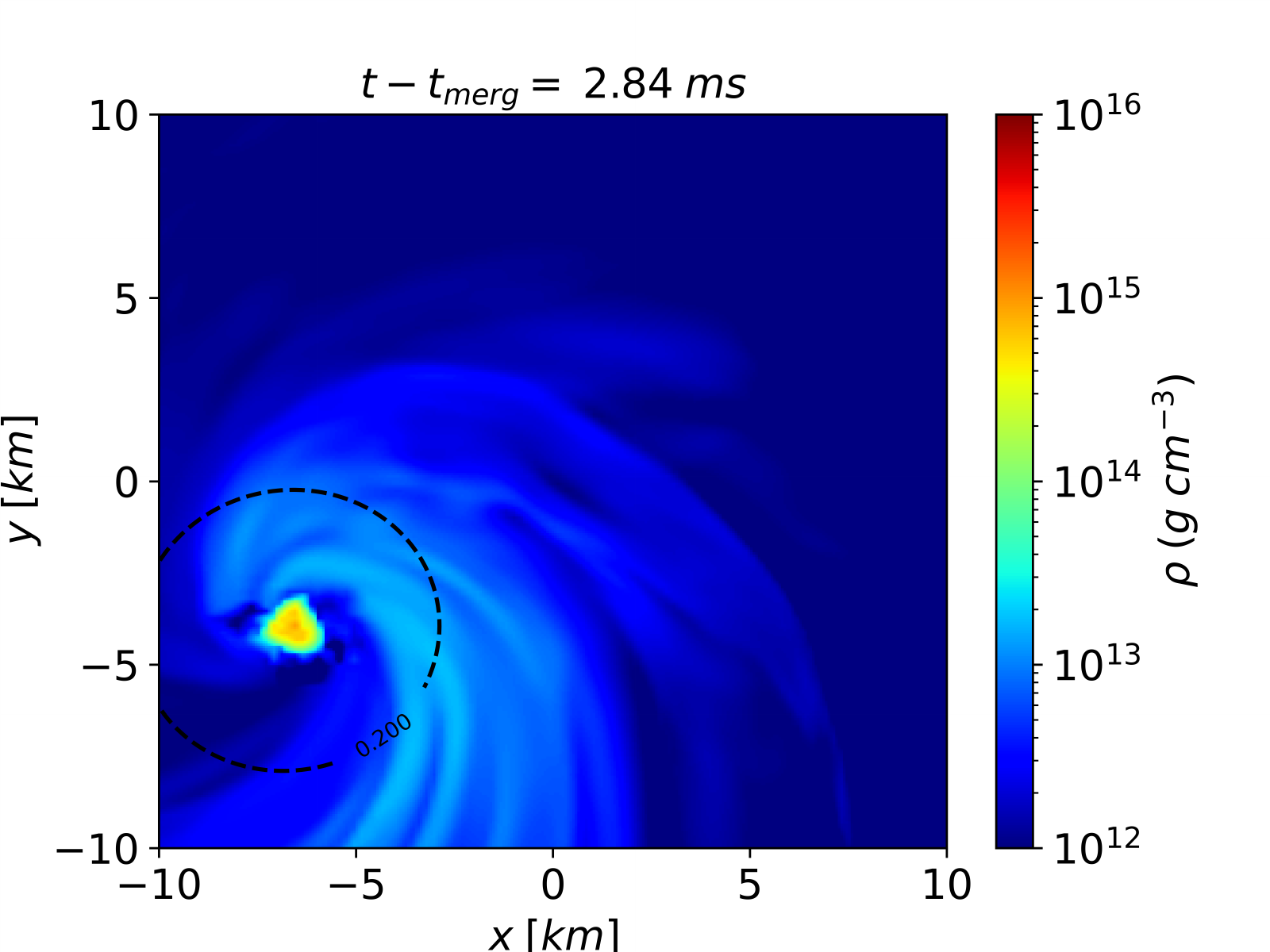}\\
&\IncG[trim={1cm 0 1cm 0}, width=.35\textwidth,height=.3\textwidth]{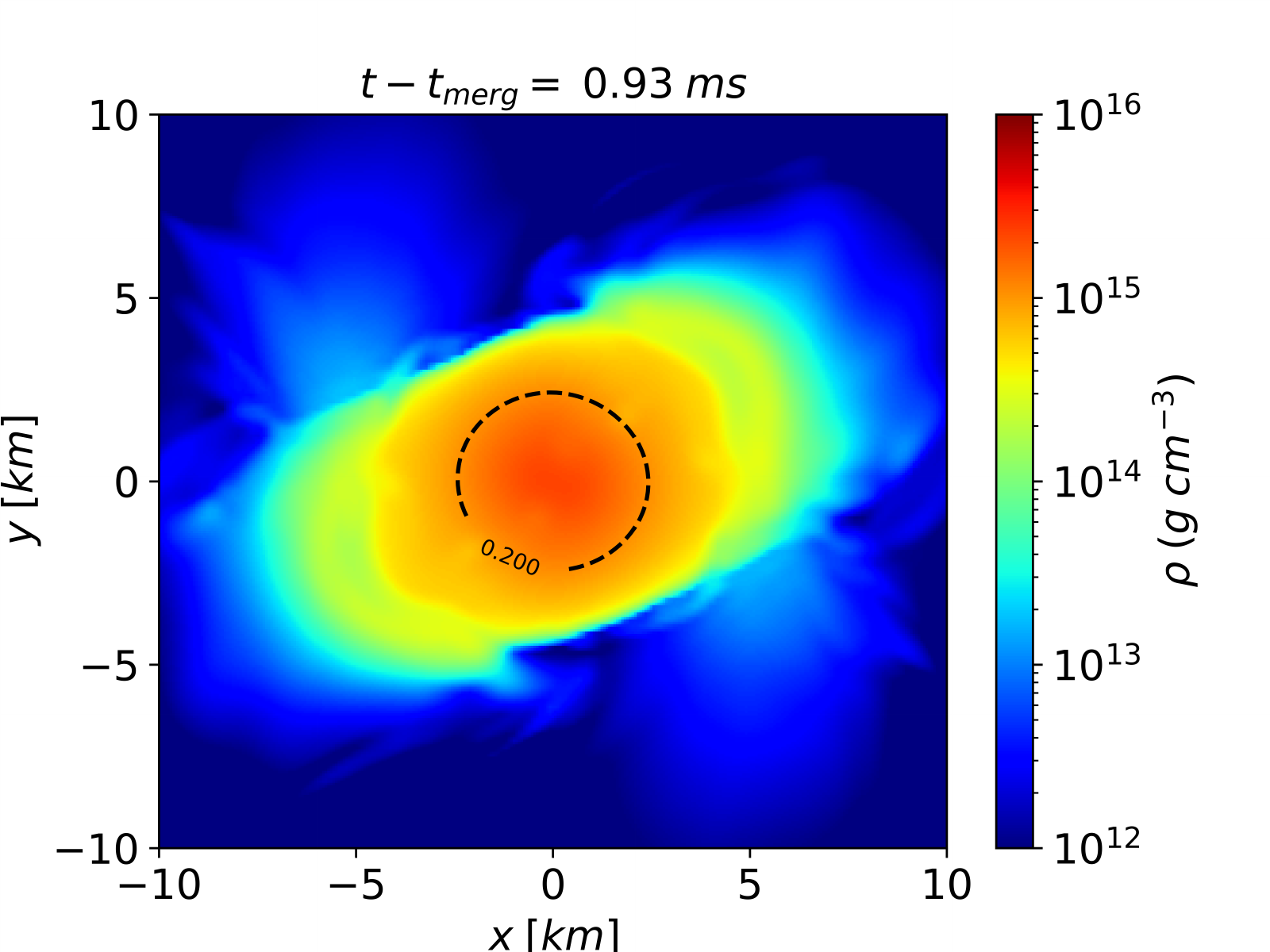}
&\IncG[trim={1cm 0 1cm 0}, width=.35\textwidth,height=.3\textwidth]{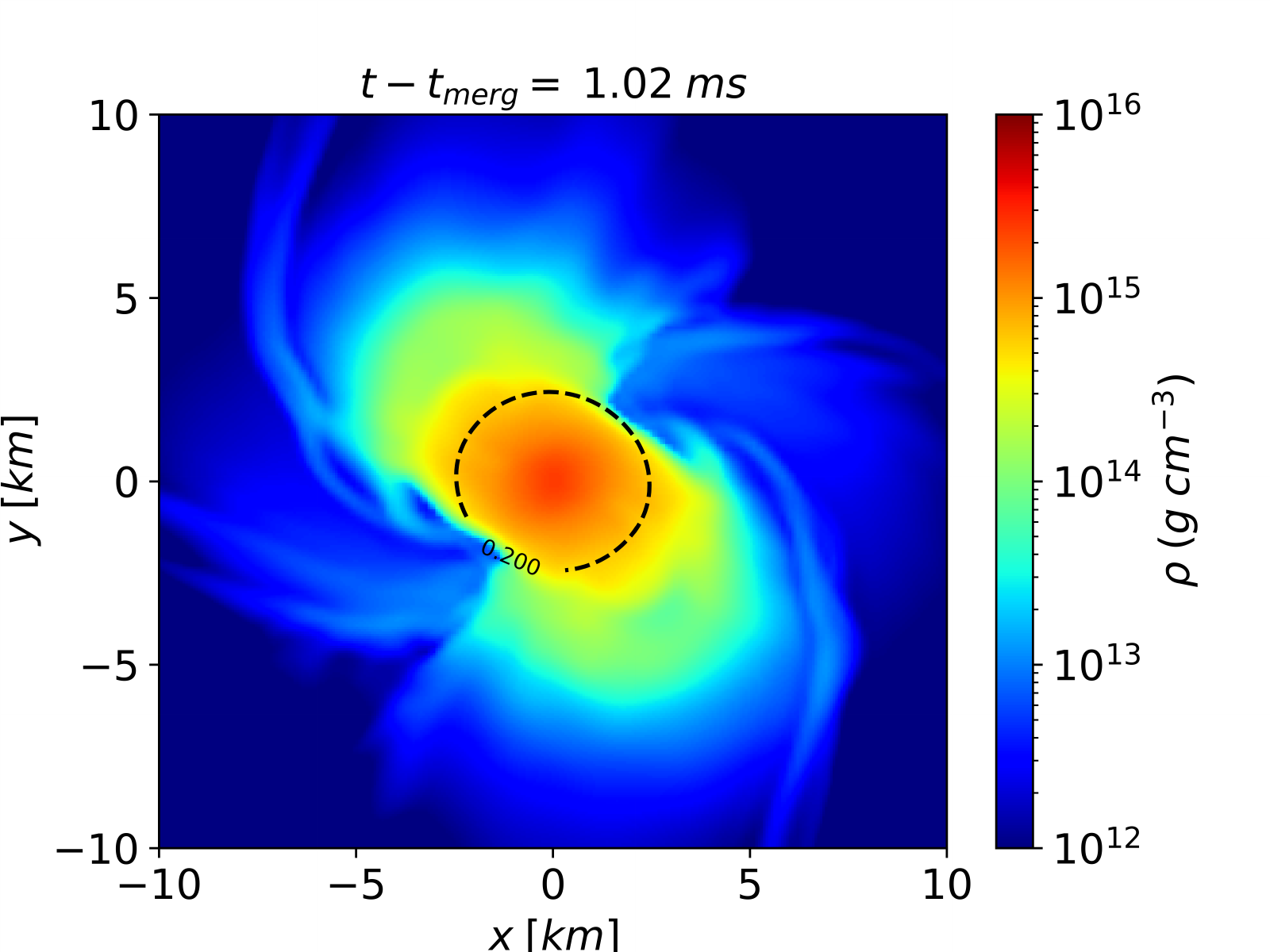}
&\IncG[trim={1cm 0 0 0}, width=.35\textwidth,height=.3\textwidth]{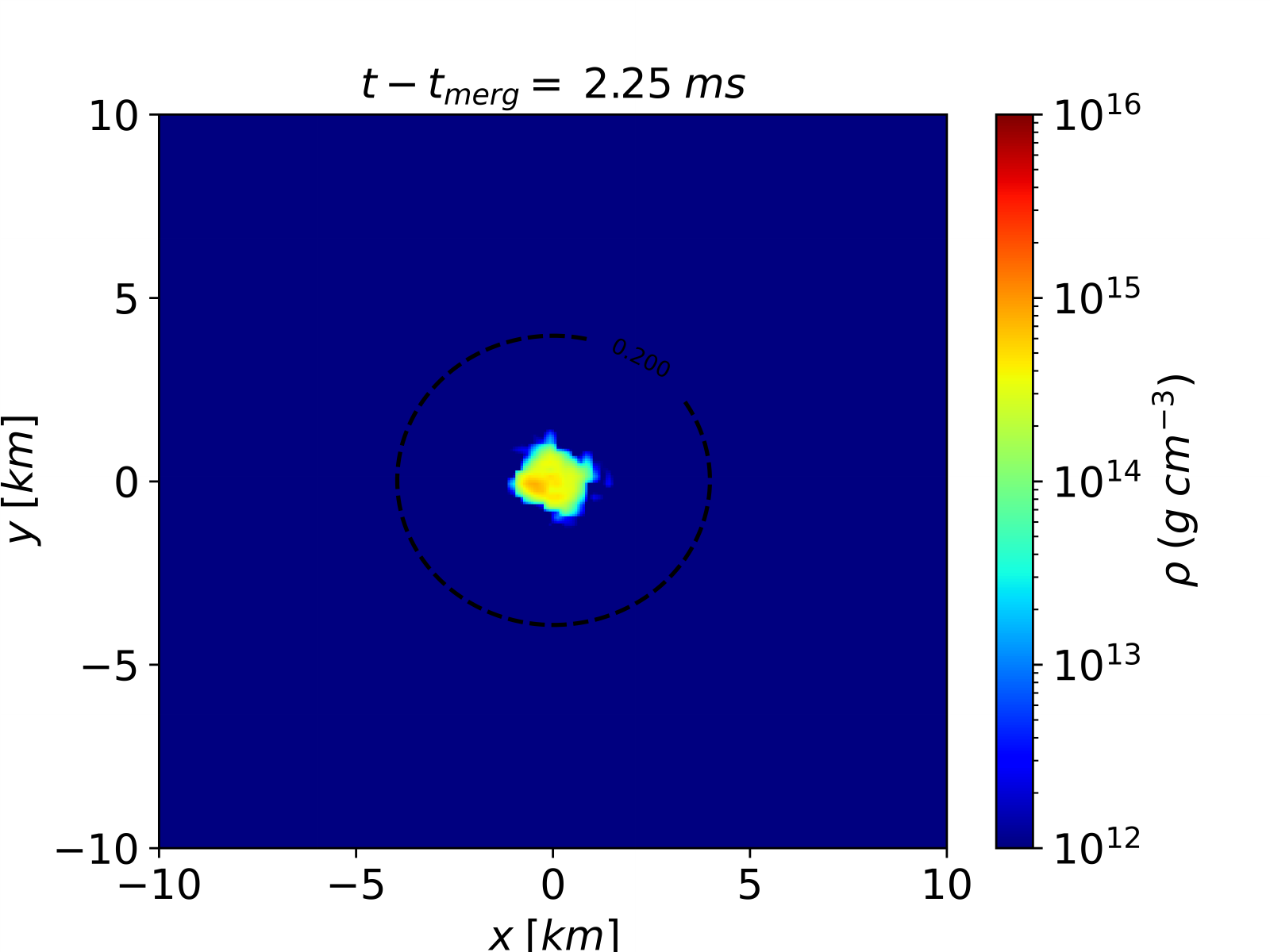}\\
\end{tabular}
\caption{Color map of rest mass density in two-dimensional slices taken along the equatorial plane of the binaries. The contour in some of the frames corresponds to lapse equals 0.2. Each row corresponds to slices from a different binary with individual masses, from top to bottom, 1.35 $M_\odot$, 1.45 $M_\odot$ and 1.5 $M_\odot$, respectively. Each binary is created using the EOS SRO2. In the middle row, the drift in the formed BH is a gauge effect and does not correspond to actual movement.}
\label{fig:slices}
\end{figure*}

\begin{figure*}
    \includegraphics[width=\linewidth]{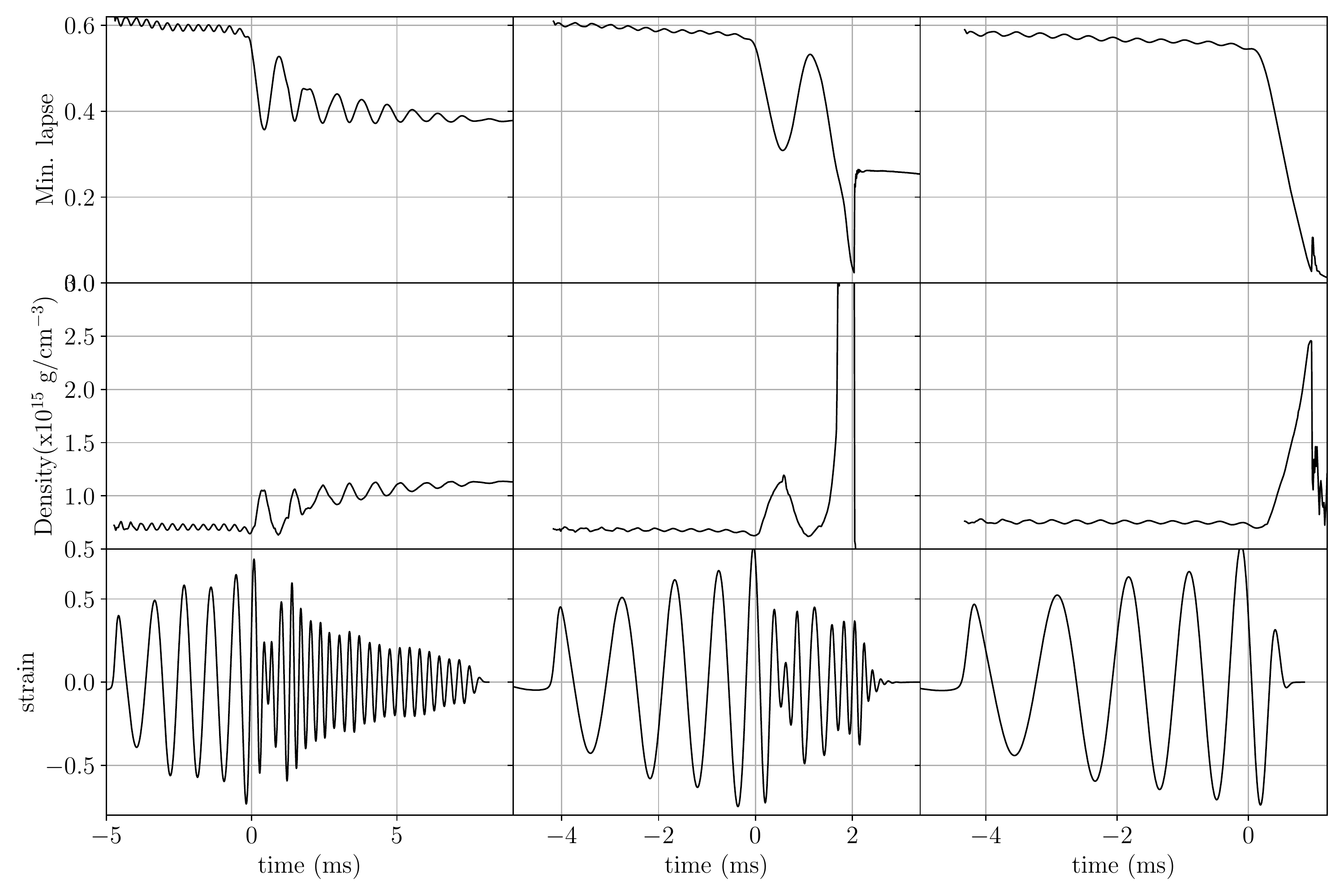}
    \caption{Examples of the minimum lapse function, maximum rest mass densities and $l=2,m=2$ mode gravitational wave strains (in geometric unit of total mass of binaries) from three different simulations (\textit{from left to right}: EOS2-1.36 $M_\odot$, EOS6-1.48 $M_\odot$ and EOS2-1.51 $M_\odot$ binary systems). The first column shows a binary resulting in a massive neutron star that does not collapse during the simulation. The binary in the second column produces a massive neutron star that rapidly collapses to a black hole after a single core bounce. The binary shown in the third column collapses immediately upon merger and is classified as prompt collapse.}
    \label{fig:1dplots}
\end{figure*}

In Fig.\ref{fig:slices}, we show slices of the rest mass density of the rest mass density on the orbital plane from three different binaries representative of typical outcomes. We chose EOS SRO2 for this demonstration. Moreover, in slices where the value of the lapse function drops below 0.2, we have included a contour indicating the spatial location where it happens. This is a rough estimate of the apparent horizon and the appearance of this contour indicates that the merger remnant has collapsed and formed a black hole \citep{Bernuzzi:2020txg}. 

The first row involving stars of gravitational mass 1.35 M$_\odot$ each, presents a merger resulting in a long-lived merger remnant (or a massive NS). The left panel shows the moment just after the merger as the two NSs make contact. The second panel demonstrates a merger remnant with a double core structure surrounded by a common envelope. The third and final panel shows a single rotating neutron star with an oscillating core, losing energy and angular momentum through the emission of gravitational waves to achieve greater stability more than 10 ms after the merger. For the minimum lapse, maximum of the rest mass density and gravitational wave strain as a function of time, see the first column of Fig. \ref{fig:1dplots}.

The second row shows a binary of component mass 1.45 M$_\odot$ which undergoes delayed collapse. Like in the previous case, the left panel represents the remnant just after merger. Plumes of ejecta can be seen emerging as the remnant spins with the residual angular momentum of the merged binary. In the middle panel, the core of the remnant has already collapsed as indicated by the lapse decreasing below 0.2 and ejection of the outer layers of matter continues to take place. The right panel shows the remnant after most of the matter has either been ejected or accreted onto the black hole. A small amount of matter can be seen forming a disk around the black hole. Time series of minimum lapse, maximum rest mass density and gravitational wave strain for this binary can be found in the second column of Fig. \ref{fig:1dplots}.

The third row corresponds to a binary undergoing prompt collapse with a component mass of 1.5 M$_\odot$. The first panel, which shows the remnant $\sim$ 1 ms after merger, indicates that the core has collapsed as the lapse has fallen below 0.2 at which point the two neutron stars are no longer distinguishable. The second frame sees the remnant shrink in size as more and more matter falls into the black hole or gets ejected. In the final frame, only the black hole remains with little to no trace of matter surrounding it. See the third column of Fig. \ref{fig:1dplots} for quantities corresponding to this binary. 

We calculate $k_{th}, \Lambda_{th}$ and report them in Tab.~\ref{tab:main} along with properties of the NSs corresponding to the EOSs used in our simulations.

\subsection{Constraints on Neutron Star Radii and maximum masses}
\label{sec:constraints_radii}

\begin{figure*} 
    \centering
    \includegraphics[width=\linewidth]{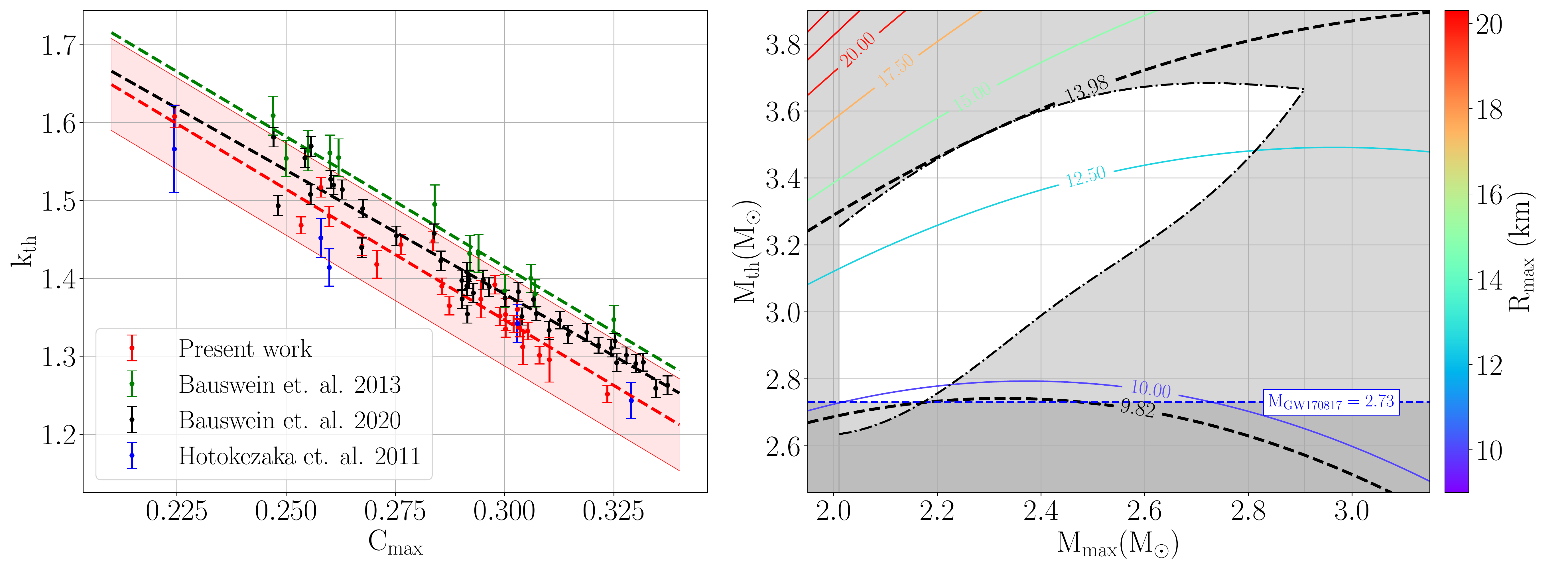}
    \caption{\textit{Left panel:} Plot of $k_\text{th}$ vs. $C_\text{max}$ from present and previous works~\cite{Hotokezaka:2011dh,Bauswein:2013jpa,Bauswein:2020xlt}. Fits are constructed using our data and are shown in combination with the data of \citet{Hotokezaka:2011dh},  \citet{Bauswein:2013jpa} and \citet{Bauswein:2020xlt}. The weighted linear regression results take into account the uncertainty in $k_{th}$. The shaded region represents uncertainties in the intercept. \textit{Right panel:} Constraints on the $R_{max}$, \mmax{} and \mth{} obtained using the correlation in left panel, PWP phenomenological constraints in combination with the observational lower limit on the maximum mass of nonrotating neutron stars and total mass of the event GW170817 as the lowest limit for prompt collapse. 
    }
    \label{fig:kth_rmax_mth}
\end{figure*}

Following the literature \citep{Bauswein:2013jpa,Bauswein:2017vtn,Bauswein:2020aag}, we assume $M_\text{th}$ for each EOS to be directly proportional to the maximum mass ($M_\text{max}$) for a nonrotating NS predicted by that EOS. Relations between \mth{}, \mmax{} and \cmax{} are given as: 
\begin{subequations}
    \begin{equation}
    \label{eqn:defmth}
        M_\text{th} = k_\text{th}M_\text{max}  
    \end{equation}
    \begin{equation}
    \label{eqn:defkth_cmax}
    	k_{th} = a C_{max} + b
    \end{equation}
    \begin{equation}
    \label{eqn:Mth_cmax}
    	M_\text{th} = (a C_{max} + b)M_\text{max}
    \end{equation}
\end{subequations}
where $a$ and $b$ are coefficients of linear fits obtained using weighted least squares method. Re-arranging the linear fit equation for $k_{th}$ vs $C_{max}$, we obtain the following empirical relations:

\begin{subequations}
	\begin{equation}
	\label{eqn:Rmax_Mth_mmax} 
        R_{max}  = \frac{G}{c^2} \left[ \frac{aM_{max}^2}{M_{th}-bM_{max}}\right]   
    \end{equation}
    \begin{equation}
	\label{eqn:Mth_Rmax_mmax}
        M_{th} = \left[ a\frac{G M_{max}}{R_{max}c^2} + b  \right] M_{max}  
    \end{equation}
\end{subequations}
where $a,b$ are fitting coefficients in Eq.~\eqref{eqn:defkth_cmax}. We report these fits in Tab.~\ref{tab:fits} for our data as well as various data sets in the literature. Using the Eq.~\eqref{eqn:Rmax_Mth_mmax}, we plot the constant $R_{max}$ contours in Fig.~\ref{fig:kth_rmax_mth}. 

We find agreement with the claim that $k_{th}$ is directly proportional to $C_{max}$ and modified compactnesses ($C^*_{1.4},C^*_{1.6}$) which are defined as follows: 
\begin{subequations}
	\begin{equation}
	\label{eqn:modcs14tar}
		C^*_{1.4}:=\frac{GM_\text{max}}{c^2 R_\text{1.4}}
	\end{equation}
	
	\begin{equation}
	\label{eqn:modcs16tar}
		C^*_{1.6}:=\frac{GM_\text{max}}{c^2 R_\text{1.6}}
	\end{equation}
\end{subequations} where $R_{1.4}$ and $R_{1.6}$ are radii of a 1.4 $M_\odot$ and a 1.6 $M_\odot$ NS.
We find the R-squared goodness of fit for $k_{th}-C_{max}$, $k_{th}-C^*_{1.6}$ and $k_{th}-C^*_{1.4}$ to be 0.93, 0.90 and 0.86, respectively.
From the left panel of Fig. \ref{fig:kth_rmax_mth} and Tab.~\ref{tab:fits}, we can see that our data agrees closely with that of \citet{Hotokezaka:2011dh} which uses an independent \revf{$3+1$ numerical relativity} implementation for solving Einstein's equations. On the other hand, we see a systematic deviation from the results of ~\citet{Bauswein:2013jpa} which uses \revf{an implementation of conformally flat approximation. However, the improved version of their viscosity treatment gives consistent result with our fitting coefficients \citep{Bauswein:2020xlt}.}

\begin{figure*} 
    \centering
    \includegraphics[width=\linewidth]{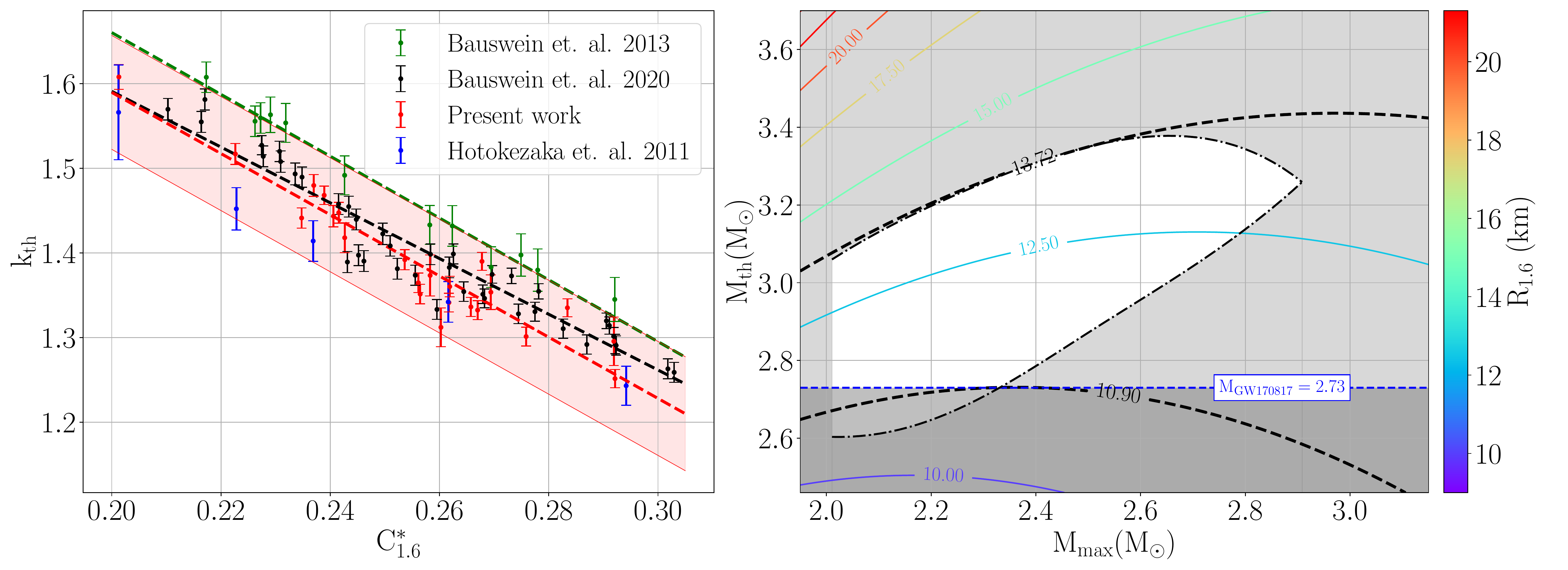}
    \caption{\textit{Left panel:} Plot of $k_\text{th}$ vs. $C^*_{1.6}$ from present and previous works~\cite{Hotokezaka:2011dh,Bauswein:2013jpa,Bauswein:2020xlt}. Fits are constructed using our data and are shown in combination with the data of \citet{Hotokezaka:2011dh}, \citet{Bauswein:2013jpa}  and \citet{Bauswein:2020xlt}. The weighted linear regression results take into account the uncertainty in $k_{th}$. The shaded region represents uncertainties in the intercept. \textit{Right panel:} Constraints on the $R_{1.6}$, \mmax{} and \mth{} obtained using the correlation in left panel, PWP phenomenological constraints obtained using the correlation in the left panel, PWP phenomenological constraints in combination with the observational lower limit on the maximum mass of nonrotating neutron stars and total mass of the event GW170817 as the lowest limit for prompt collapse.
    }
    \label{fig:kth_r16_mth}
\end{figure*}

\begin{figure*} 
    \centering
    \includegraphics[width=\linewidth]{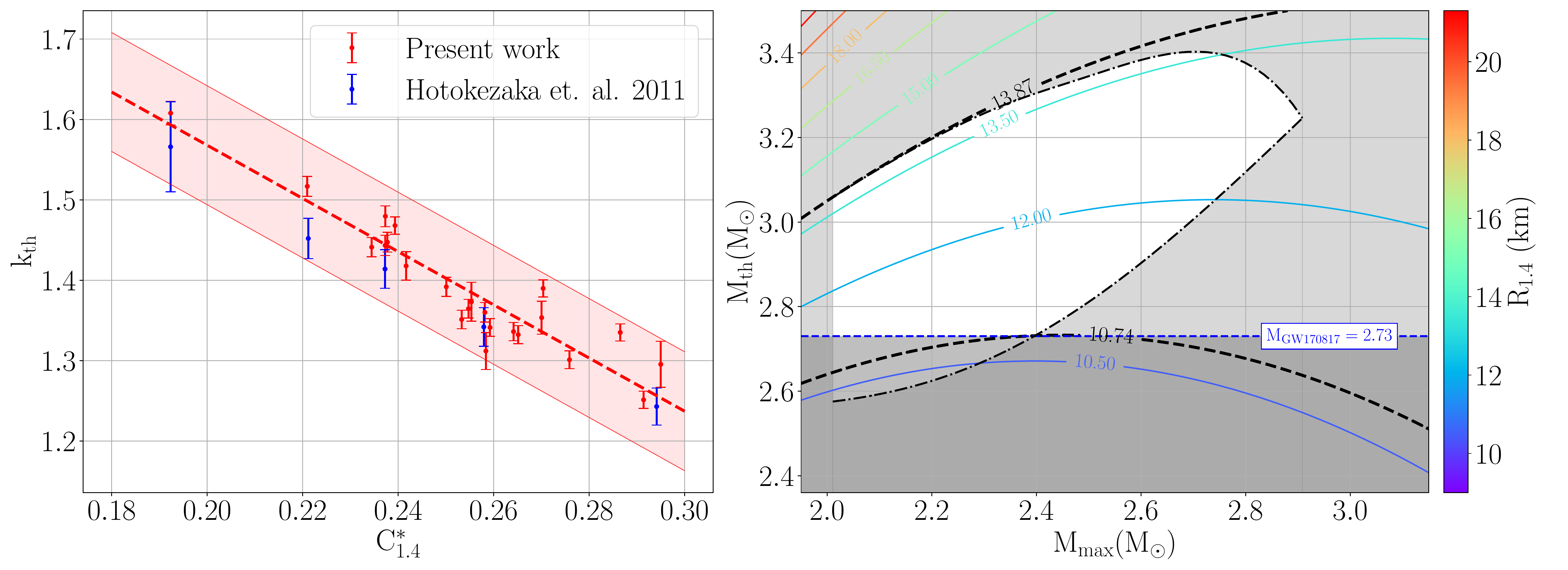}
    \caption{\textit{Left panel:} Plot of $k_\text{th}$ vs. $C^*_{1.4}$ from present and previous works~\cite{Hotokezaka:2011dh}. Fits are constructed using our data in combination with the data of \citet{Hotokezaka:2011dh}. The weighted linear regression results take into account the uncertainty in $k_{th}$. The shaded region represents uncertainties in the intercept. \textit{Right panel:} Constraints on the $R_{1.4}$, \mmax{} and \mth{} obtained using the correlation in left panel, PWP phenomenological constraints obtained using the correlation in the left panel, PWP phenomenological constraints in combination with the observational lower limit on the maximum mass of nonrotating neutron stars and total mass of the event GW170817 as the lowest limit for prompt collapse.}
    \label{fig:kth_r14_mth}
\end{figure*}

Using the linear correlation between $k_{th}$ and $C_{max}$ discussed above and the phenomenological constraints on the absolute maximum value of $C_{max}$, it is possible to derive a lower bound for $R_{max}$ \cite{Bauswein:2017vtn, Koppel:2019pys}. We recompute this constraint using our fit of $k_{th}$ vs $C_{max}$. 

Independent phenomenological constraints on upper and lower limits of $R_{max}$ have also been obtained using our data \cite{Godzieba:2020bbz,Godzieba:2021vnz,Godzieba:2020tjn} which depend on the maximum masses in the range 1.97 $M_{\odot}$ - 2.9 $M_{\odot}$ as shown in Fig. \ref{fig:gz_Cminmax}. We use 2 million piecewise polytropic (PWP) EOSs to construct the key properties such as mass, radius and tidal deformability of 1.4 $M_\odot$, 1.6 $M_{\odot}$ and maximum mass NSs (see appendix \ref{app:gzlimits} for details). Our data shows the maximum value of \mmax{} to be 2.9 $M_\odot$ which rules out the region right of this value in Fig. \ref{fig:kth_rmax_mth}. The minimum value of \mmax{} is determined by the maximum mass pulsar observed till date ($\sim 2.01~M_\odot$) which excludes the region left of it. We use Eq.~\eqref{eqn:Mth_Rmax_mmax} to calculate the limits of $M_{\rm{th}}$ at each value of \mmax{} corresponding to minimum and maximum value of $R_{max}$ taken from Fig.~\ref{fig:gz_Cminmax}. The computed threshold mass limits (corresponding to $R_{max}$ limits) are plotted as dot-dashed black lines on the right panel in Fig.~\ref{fig:kth_rmax_mth} ruling out the region above the upper and below the lower line. The horizontal line corresponding to the mass of GW170817 rules out the region below it due to its identification as a delayed collapse event.  The contour of constant R$_{\rm{max}}$ passing through the lowest point of unshaded region provides the minimum allowed value of R$_{\rm{max}}$ (see Fig. \ref{fig:kth_rmax_mth}). 

Similarly, using our data of simulated PWP EOSs, we also plot the \mmax-dependent minimum and maximum values of $R_{1.4}$ and $R_{1.6}$ as shown in Fig.~\ref{fig:gz_Cs16minmax} and \ref{fig:gz_Cs14minmax}. We define linear relations between k$_{th}$ and the modified compactness parameters, $C^*_{1.6}$ and $C^*_{1.4}$ similar to Eq.~\eqref{eqn:defkth_cmax}. We obtain expressions for R$_{1.4}$ and R$_{1.6}$ similar to Eq.~\eqref{eqn:Rmax_Mth_mmax} as well as for M$_{th}$ as a function of R$_{1.4}$ and R$_{1.6}$ similar to Eq.~\eqref{eqn:Mth_Rmax_mmax}. Our newly found correlation between $k_{th}$ and modified compactness, $C^*_{1.4}$ (see Fig.~\ref{fig:kth_r14_mth}) provides a constraint on R$_{1.4}$ which is a known constraining factor for pressure at two times saturation density \cite{Lattimer:2015nhk}.

The same procedure as described above is used to plot contours of constant $R_{1.6}$ in Fig.~\ref{fig:kth_r16_mth} and of constant $R_{1.4}$ in Fig.~\ref{fig:kth_r14_mth}.  Constraints for $R_{1.6}$ and $R_{1.4}$ and excluded regions for \mth{} and \mmax{} are obtained analogous to $R_{max}$ as shown in Figs.~\ref{fig:kth_r16_mth} and \ref{fig:kth_r14_mth}, respectively. We report the constraints on minimum values of $R_{1.6}$ and $R_{1.4}$ in Tab.~\ref{tab:fits} with the uncertainty that takes into account the uncertainties in the mass of GW170817 as well as our derived value of intercepts for the linear fits.

\revf{Extending the method from the previous works \citep{Bauswein:2017vtn,Bauswein:2020aag,Bauswein:2020xlt}, we find that the constraints on the threshold masses} as shown in the figures above provide a method to constrain the lower and upper limits on the maximum mass of non-rotating NS as follows. Due to lower (upper) limit on $R_{max}$, we obtain an lower (upper) limit on $M_{\rm{th}}$ for a given value of $M_{max}$ using Eq.~\eqref{eqn:Mth_Rmax_mmax} as shown in Fig.~\ref{fig:kth_rmax_mth}. The dependence of $M_{\rm{th}}$ bounds on $M_{\rm{max}}$  implies constraints on the maximum mass of NS \textit{if we infer an unambiguous prompt collapse or delayed collapse in a future BNS detection}. This is so because observation of a prompt collapse event with total mass ($M_{\rm{total, Prompt}}$) less than $3.64~M_{\odot}$ (Tab.~\ref{tab:constraints}) will exclude the region above the horizontal line corresponding to that mass. If observed in future GW detections, and with $M_{total}$ greater than this value, we will obtain a constraint on the upper bound of maximum mass better than the one obtained using PWP phenomenological constraints ($\approx 2.9 \rm{M_{\odot}}$). The constraining upper value of maximum mass will be the intersection point of the horizontal line at $M_{\rm{total, Prompt}}$ and the upper bounding curve of $M_{\rm{th}}$ in Fig. \ref{fig:kth_rmax_mth}. One can also deduce another constraining fact from these figures that \textit{if we observe an event with delayed collapse, like GW170817, with total mass ($M_{\rm{total, Delayed}}$) greater than $3.25~M_{\odot}$}, we can constrain the lower limit of maximum masses better than the current best estimate from pulsar observations. The constraint on the lower limit of the maximum mass will be the intersection point of the horizontal line corresponding to $M_{\rm{total, Delayed}}$ and the lower bounding curve of $M_{\rm{th}}$ in Fig. \ref{fig:kth_rmax_mth}.

The lower and upper constraints on $M_{\rm{max}}$ as well as maximum value of threshold masses corresponding to the correlations of $k_{\rm{th}}-C^*_{1.6}$ and $k{_{\rm{th}}}-C^*_{1.4}$ can be obtained using a similar procedure. The critical values of $M_{\rm{total, Prompt}}$ and $M_{\rm{total, Delayed}}$ from corresponding correlations as shown in Fig.~\ref{fig:kth_rmax_mth}, \ref{fig:kth_r14_mth} and \ref{fig:kth_r16_mth} are quoted in Tab.~\ref{tab:constraints}.

\subsection{Implications on Tidal Deformability}

\begin{figure*} 
    \centering
    \hspace*{-1cm}\includegraphics[width=1.2\linewidth]{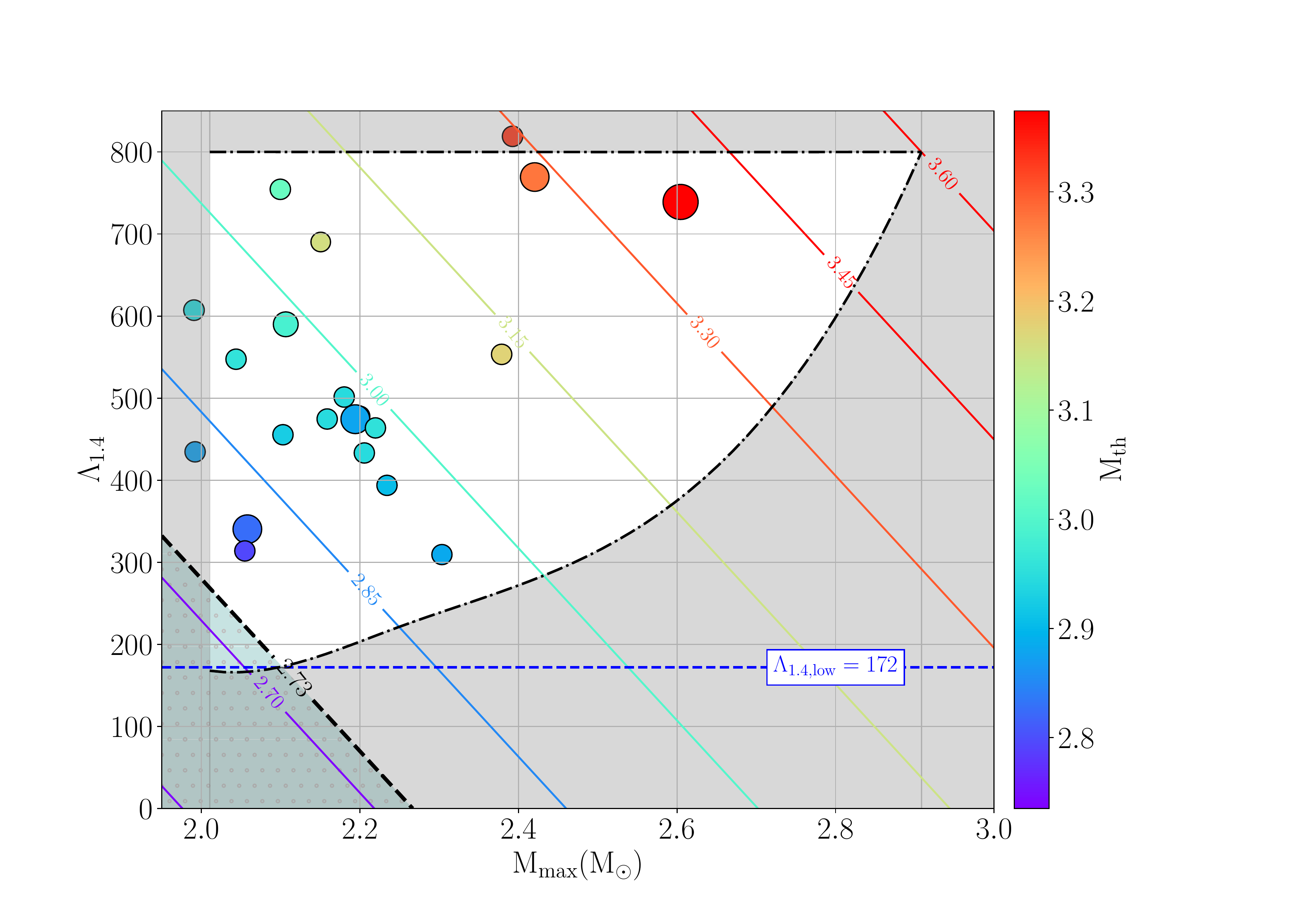}
    \caption{Data (colored circles) and linear relationships obtained from the fit in Eq.~\ref{eqn:Mthlam14mmax} relating \mth{} to \mmax{} and $\Lambda_{1.4}$ (solid contour lines). Black dot-dashed curves shows the most conservative upper and lower constraint on $\Lambda_{1.4}$ from \cite{Godzieba:2020bbz} (see also values in  \citep{TheLIGOScientific:2017qsa} $190^{+390}_{-120}$). Please note the two contours in thick dashed black lines; for 2.73 M$_\odot$ corresponding to the total mass of GW170817 ($M=2.73^{+0.04}_{-0.01}$, \cite{TheLIGOScientific:2017qsa}) and for the maximum threshold mass to be equal to 3.62 M$_\odot$. We report the upper limit of threshold mass below which we can constrain the upper limit of NS maximum mass.}
    \label{fig:Mth_Mmax_lam14}
\end{figure*}

 Following \citet{Bauswein:2020aag}, we fit the threshold mass as a bilinear function of \mmax{} and the tidal deformability of the 1.4 M$_{\odot}$ NS ($\Lambda_{1.4}$). 

\begin{equation}
    M_\text{th} (\Lambda_{1.4},M_\text{max}) = s_0 M_\text{max} + s_2 \Lambda_{1.4} + s_3
    \label{eqn:Mthlam14mmax}
\end{equation}
where the fitting coefficients obtained using our data are $s_0=0.62 \pm 0.05, s_1=(5.83 \pm 0.56) \times 10^{-4}$, and $s_3=1.33 \pm 0.11$.
We propose to combine the approach of \citet{Bauswein:2020aag} along with PWP constraints to find the lower limit of $\Lambda_{1.4}$ as follows. In addition to the data points (as colored circles) and contours of the constant M$_{\rm{th}}$, we plot several observational and phenomenological constraints in Fig.~\ref{fig:Mth_Mmax_lam14}. The lower limit of NS maximum mass from pulsar observation ($M_{\rm{max}}>2.01 M_\odot$) excludes the region left of this value while the upper limit of $\Lambda_{1.4}$ from the observation of GW170817 ($\Lambda_{1.4}<800$) eliminates the region above the horizontal line corresponding to this value. We again use our PWP phenomenological data to plot \mmax{}-dependent constraint on the lower limit of $\Lambda_{1.4}$ (lower dot-dashed black curve in Fig.~\ref{fig:Mth_Mmax_lam14}), ruling out the region below this curve. The lowest point of the allowed (unshaded) region gives the lower limit on $\Lambda_{1.4}$ ($=172$) which corresponds to the intersection point of the slanted line corresponding to $M_{th}=2.73~M_{odot}$ (total mass of GW170817) and the lower bounding curve of $\Lambda_{1.4}$.

\begin{table*}
\centering
\caption{\label{tab:fits} Fitting coefficients in Eq.~\eqref{eqn:defkth_cmax} and radius constraints obtained from our data as well as from literature. The fitting coefficients and their uncertainties are reported using weighted least square method. The uncertainties in radii comes from the uncertainties in the mass of GW170817 as well as in the values of fitting coefficients. Please note that the upper limits on radii are purely PWP phenomenological constraints.}

\begin{tabular}{c | c | c | c | c | c | c | c | c }
\hline
dataset  &  a   &   b  &  min(R$_{\text{max}}$) & max(R$_{\text{max}}$)  &  min(R$_{\text{1.6}}$) & max(R$_{\text{1.6}}$) &  min(R$_{1.4}$) & max(R$_{1.4}$) \\
  &  &  & (km) & (km) & (km) & (km)  & (km) & (km) \\
\hline
\citet{Bauswein:2013jpa}  & -3.342 & 2.42 & - & - & - & - & - & - \\
\rule{0pt}{4ex} \citet{Bauswein:2017vtn}  & -3.38 & 2.43 & 9.26$^{+0.17}_{-0.03}$ & 10.30 $^{+0.15}_{-0.03}$ & -  & - & - & - \\
\rule{0pt}{4ex}  \textbf{Current work}   & $-3.36 \pm 0.20$ & $2.35 \pm 0.06$ & 9.81$^{+1.20}_{-1.09}$  & 13.98 & 10.90$^{+1.85}_{-1.42}$ & 13.72 & 10.74$^{+1.86}_{-1.61}$ & 13.87\\
\hline
\end{tabular}
\end{table*}

\begin{figure} 
    \centering
    \includegraphics[width=\linewidth]{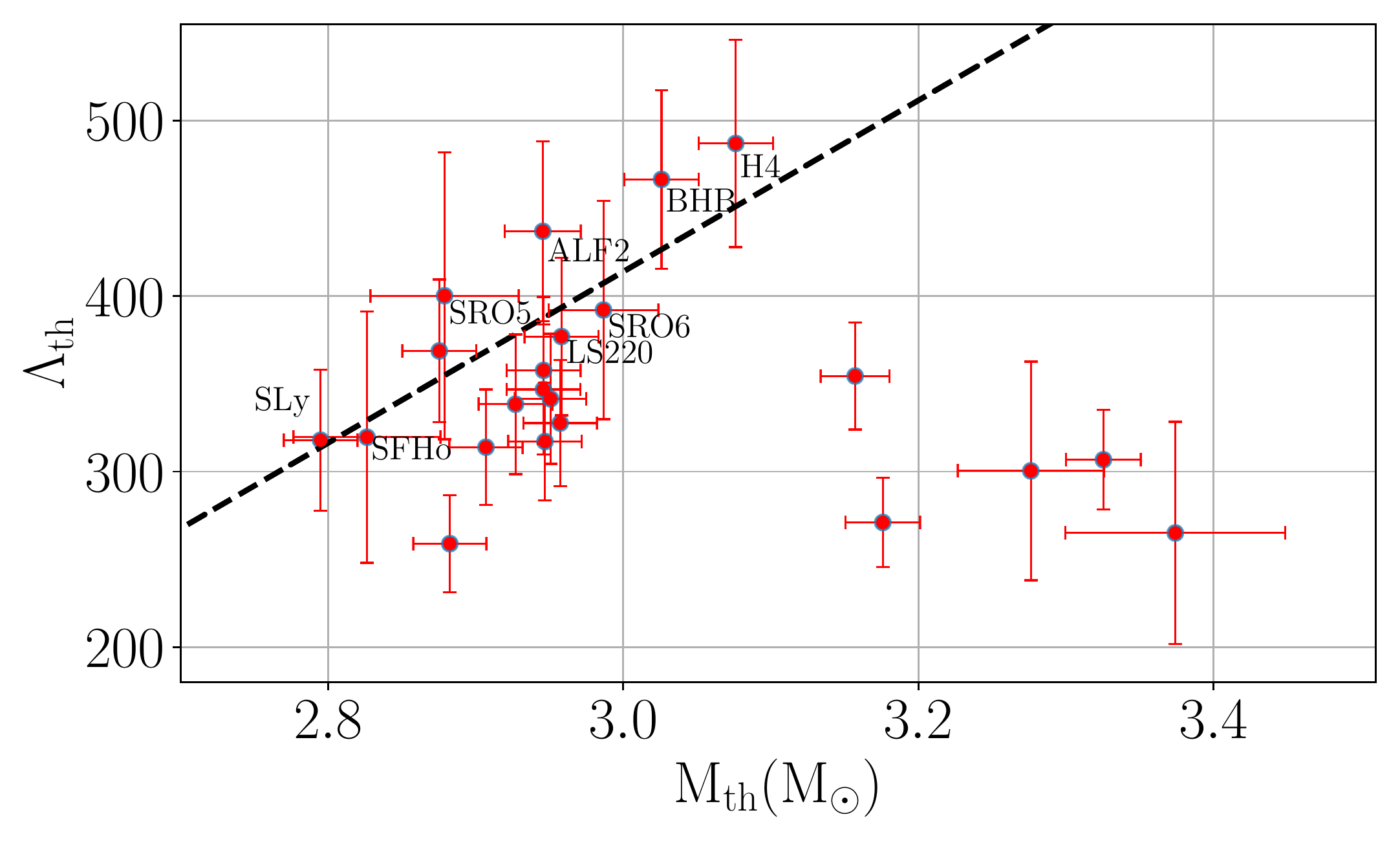}
    \caption{Plot of $\Lambda_\text{th}$ vs. $M_\text{th}$. We find one nucleonic (\textbf{SRO5}) and two non-nucleonic EOS (\textbf{BHB, H4}) whose values lies in the forbidden zone (above dashed black line as proposed by \citep{Bauswein:2020xlt}) for nucleonic EOS but, with their error bars extending below the critical line. There are few other EOSs such as \textbf{SFHo, SLy, SRO6 and LS220} whose values lie below the critical line but with error bars extending above it. The EOS, H3 lies above the critical line and outside the plot which we have removed for clarity.
    }
    \label{fig:lamth_mth}
\end{figure}

We also report the distribution of our data points on the $\Lambda_{th}$-$M_{th}$ plane where $\Lambda_{th}$ is the tidal deformability of a NS with mass equal to half of $M_{th}$ (see Fig.~\ref{fig:lamth_mth}). In agreement with \citet{Bauswein:2020xlt}, we find that most of our nucleonic EOSs lie below the critical line in this plane. However, we find that data points of one nucleonic (SRO5) and two non-nucleonic EOS (BHB, H4) lie in the forbidden region with their error bars extending below the critical line. Similarly, error bars of some nucleonic EOSs which lie below the critical line extend above it; these are SFHo, SLy, SRO6 and LS220. Further high resolution runs are required to resolve these issues and to correctly assess their claim.

\section{Conclusions and Discussions}
\label{sec:conclusions}
In this work, we present a survey of binary neutron star merger outcomes for 23 different EOS models with various assumptions about the nuclear matter. We perform numerical relativity simulations of equal mass binary NS coalescences with varying total masses and classify the outcomes as either prompt or delayed collapse. 
We confirm the correlations claimed in the literature and report a different correlation coefficients for k$_{\rm{th}}$-C$^*_{\rm{max}}$ and k$_{\rm{th}}$-C$^*_{\rm{1.6}}$ \cite{Bauswein:2013jpa,Bauswein:2017vtn} \revf{although similar to \citet{Bauswein:2020xlt}}. The linear fit reported in \citet{Bauswein:2013jpa} and \citet{Bauswein:2017vtn} lie outside the error bars obtained in our analysis \revf{but, their updated results are more consistent with ours \citep{Bauswein:2020xlt}. With the use of numerical relativity and careful error considerations, our fitting coefficients will be best suited for any future applications of this work.} We have discovered a new correlation for k$_{\rm{th}}$-C$^*_{\rm{1.4}}$ allowing us to put a lower bound on R$_{1.4}$. We find better bounds on R$_{\rm{max}}$ ($\geq$ 9.81~km) and R$_{1.6}$ ($\geq$ 10.90~km) taking into account uncertainties in the fitting coefficients and the mass of GW170817. These values are in agreements with those found by \citet{Koppel:2019pys} ($R_{1.4} \geq 9.74$ and $R_{1.6} \geq 10.90$).

Most importantly, we introduce an \mmax-dependent condition on both upper and lower limits of compactness (equivalently radius) ($C_{max}, C^*_{1.4}, C^*_{1.6}$) to derive constraint on upper and lower limits of \mmax{}. This improves upon previous methods \citet{Bauswein:2017vtn} where they use only the absolute maximum limit on compactness. We also use lower limits on compactness, not used earlier, which allows us to put novel constraints on both upper limit of \mth{} as well as lower limit of \mmax{}.
Such critical values of total masses corresponding to different correlations allow us to put bounds on \mmax{} as reported in Tab.~\ref{tab:constraints}. In addition, we also find the lower limit on $\Lambda_{1.4}$ to be 172 deduced from the bilinear fit of \mth{} as a function of \mmax{} and $\Lambda_{1.4}$ and constraints from GW170817 (see Fig. \ref{fig:Mth_Mmax_lam14}) 
We emphasize that the phenomenological constraints on R$_{max}$, R$_{1.4}$ and R$_{1.6}$ are obtained using the formulation of EOS as a piecewise polytropic pressure-density curve and hence inherit its limitations.

\begin{table}
\centering
\caption{\label{tab:constraints} Critical values of total binary masses for future GW observations for the present methods to put constraints on \mmax{}. Note that the current minimum threshold mass for prompt collapse is set by $M_{\rm{GW170817}}$=2.73. We report the total binary masses above which a potential delayed collapse will constrain the minimum value of \mmax{} as well as total masses below which a future prompt collapse event will constrain the maximum value of \mmax.}
\begin{tabular}{c | c | p{2cm} | p{2cm}  }
\hline
Correlation  &  max(\mth )   &    M$_{\rm{total,\textbf{Delayed}}}$ to   &   M$_{\rm{total,\textbf{Prompt}}}$ to \\
    &   & constrain min($M_{\rm{max}}$) & constrain max(M$_{\rm{max}}$) \\
  &  ($M_\odot$)  & ($M_\odot$) & ($M_\odot$) \\
\hline
k$_{th}$-C$_{max}$ & 3.67 & $>$ 3.25 & $<$ 3.64 \\
k$_{th}$-C$^*_{1.4}$  & 3.38 & $>$ 3.05 & $<$ 3.20 \\
k$_{th}$-C$^*_{1.6}$  & 3.35 & $>$ 3.05 & $<$ 3.22 \\
M$_{\rm{th}}$-($\Lambda_{1.4}$,\mmax{}) & 3.62 & - & - \\ 
\hline 
\end{tabular}
\end{table}

We demonstrate that the method discussed here to derive lower and upper constraints can be applied to future GW observations and their identification as delayed and prompt collapse, respectively. GW detectors are being planned to higher sensitivity which will increase the detection range as well as the accuracy of the measurements, most notably of the total masses. In future GW detectors, it will be possible to determine the likelihood of a prompt or delayed collapse following the merger using the presence or absence, respectively, of a sharp cutoff in the GW from a BNS event along with a lack of kilonova observation. Such estimate will introduce horizontal lines on Fig.~\ref{fig:kth_rmax_mth}, \ref{fig:kth_r16_mth} and \ref{fig:kth_r14_mth} ruling out regions above or below them for a prompt or a delayed collapse, respectively. In the era of highly sensitive GW observatories, tens of thousands of BNS event are expected to be detected. We are hopeful that the methodology presented here will put a strong constraint on the maximum masses and hence on the EOS of the nuclear matter.  We note the limitations in our work due to unexplored regions of BNS parameter space due to neglected effect such as spin of component NSs. We intend to extend and improve our study in a future work. 

\begin{acknowledgments}
This research was funded by U.S. Department of Energy, Office of Science, Division of Nuclear Physics under Award Number(s) DE-SC0021177 and by the National Science Foundation under Grants No. PHY-2011725, PHY-2020275, PHY-2116686, and AST-2108467.
S.~B. acknowledges support by the EU H2020 under ERC Starting Grant, no.~BinGraSp-714626.  
NR simulations were performed on Bridges, Comet, Stampede2 (NSF XSEDE allocation TG-PHY160025), NSF/NCSA Blue Waters (NSF AWD-1811236) supercomputers. Computations for this research were also performed on the Pennsylvania State University’s Institute for Computational and Data Sciences’ Roar supercomputer.
Computations were also performed on the supercomputer SuperMUC-NG at the
        Leibniz-Rechenzentrum Munich, and on the
        national HPE Apollo Hawk at the High Performance Computing
        Center Stuttgart (HLRS).
        The authors acknowledge the Gauss Centre for Supercomputing
        e.V. (\url{www.gauss-centre.eu}) for funding this project by providing
        computing time to the GCS Supercomputer SuperMUC-NG at LRZ
        (allocation {\tt pn68wi}).
        The authors acknowledge HLRS for funding this project by providing
        access to the supercomputer HPE Apollo Hawk under the grant
        number {\tt INTRHYGUE/44215}.
Finally, computations were also performed on the supercomputer Joliot-Curie at GENCI@CEA and AP
acknowledge PRACE for awarding him access to Joliot-Curie at GENCI@CEA.

\end{acknowledgments}

\bibliography{apssamp,bns_pbh_paperpile}


\appendix 

\section{Results using data from \citep{Godzieba:2020bbz}}
\label{app:gzlimits}
\begin{figure}
    \centering
    \includegraphics[width=\linewidth]{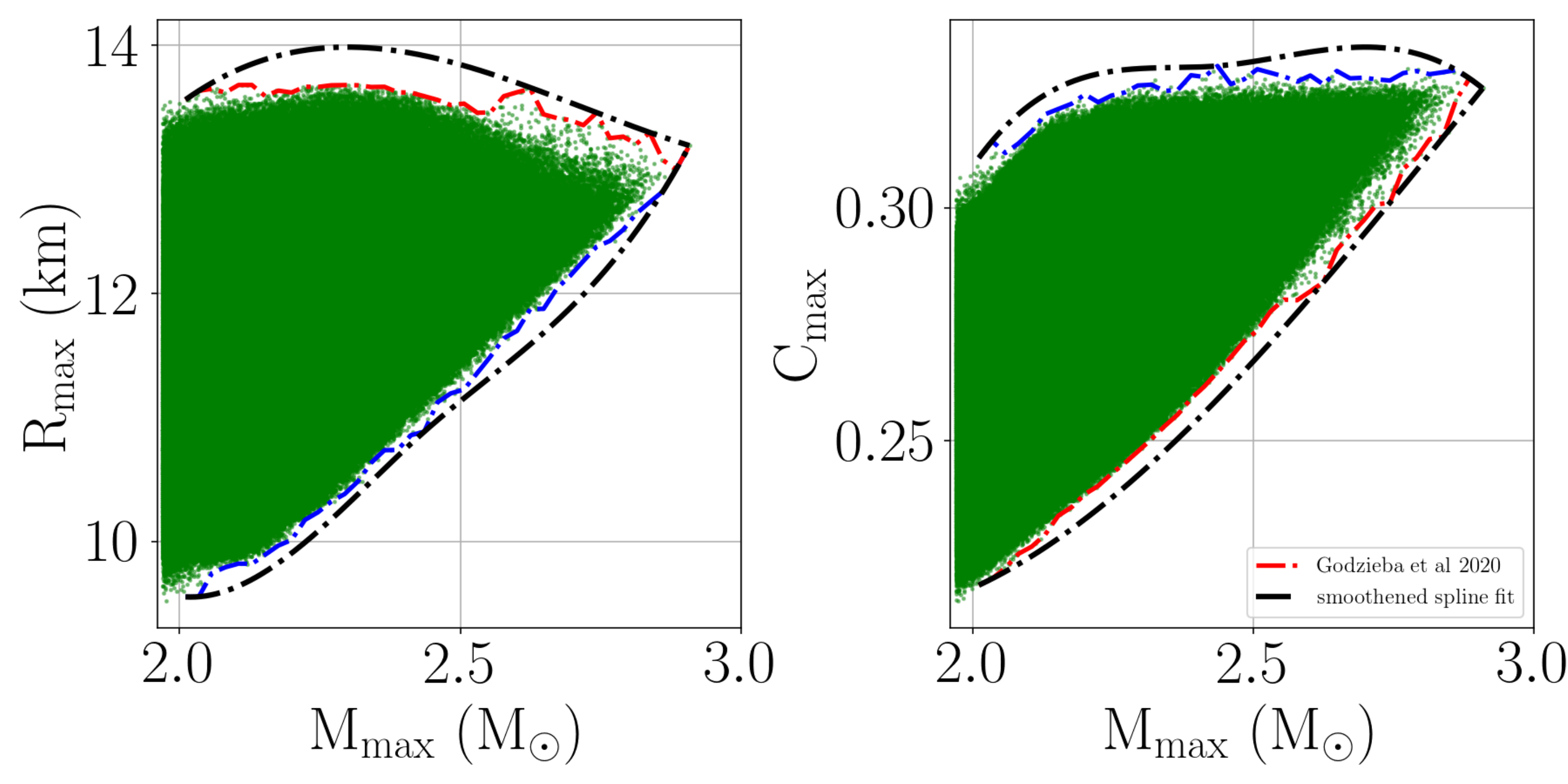}
    \caption{Maximum and Minimum compactness and corresponding radii of maximum mass NS as a function of maximum mass. The dashed black line represents a conservative bound while the red and blue dashed lines represent the actual limit of the dataset.
    }
    \label{fig:gz_Cminmax}
\end{figure}

\begin{figure}
    \centering
    \includegraphics[width=\linewidth]{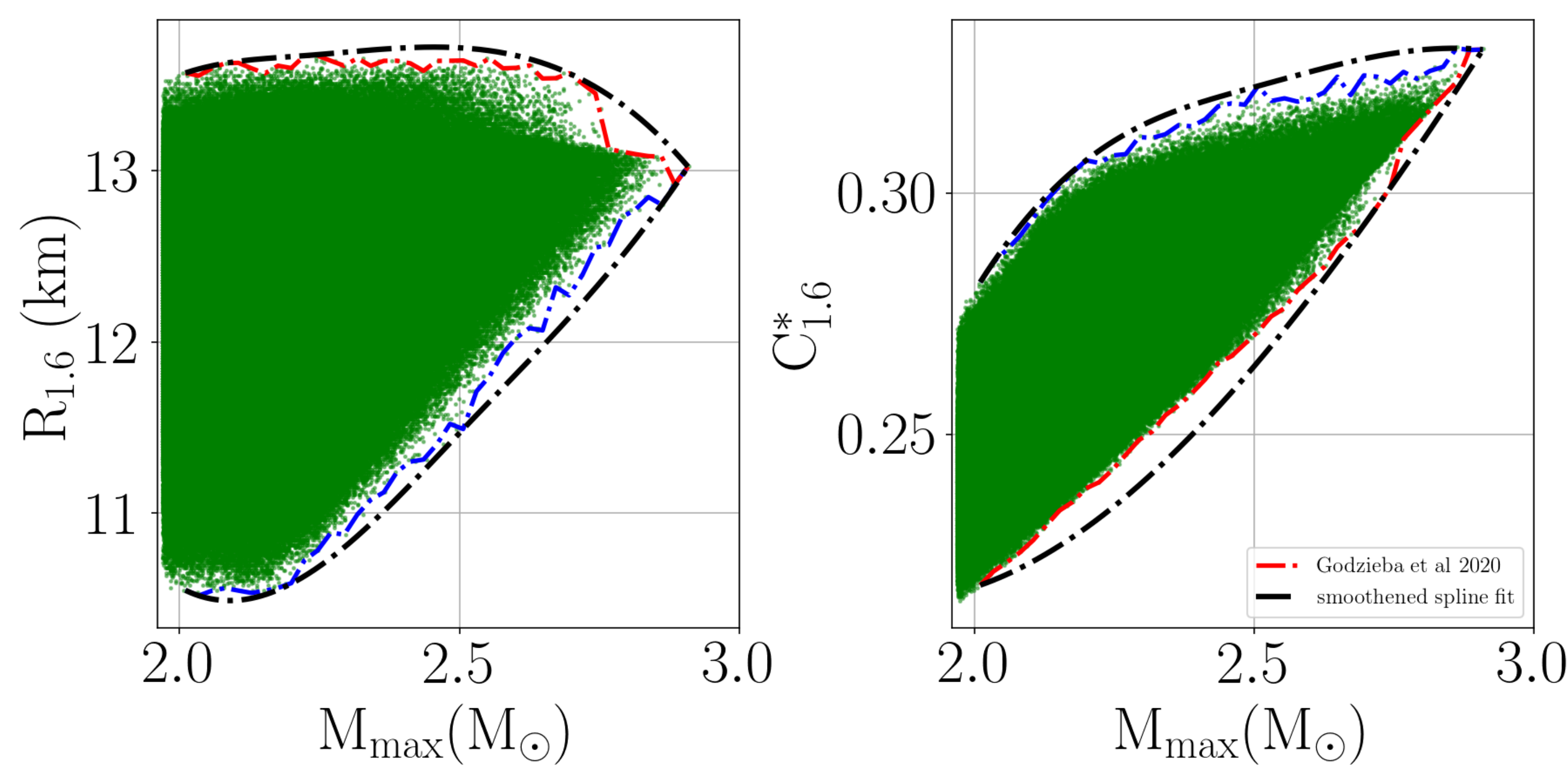}
    \caption{Maximum and Minimum modified compactness ($C^*_{1.6}$, Eq.~\eqref{eqn:modcs16tar}) and corresponding radii of a 1.6 $M_\odot$ NS as a function of maximum mass. The dashed black line represents a conservative bound while the red and blue dashed lines represent the actual limit of the dataset.
    }
    \label{fig:gz_Cs16minmax}
\end{figure}

\begin{figure}
    \centering
    \includegraphics[width=\linewidth]{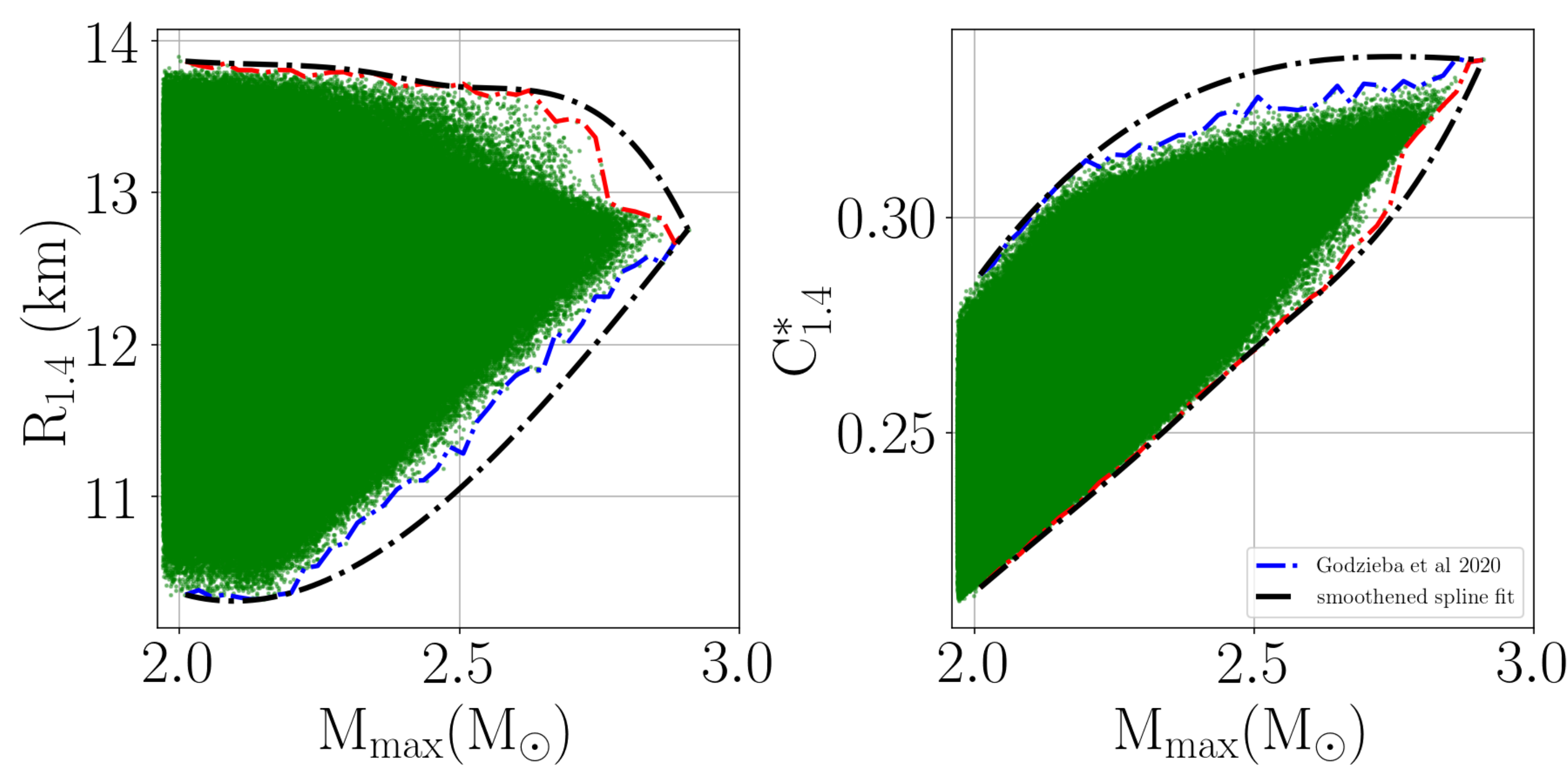}
    \caption{Maximum and Minimum modified compactness ($C^*_{1.4}$, Eq.~\eqref{eqn:modcs14tar}) and corresponding radii of a 1.4 $M_\odot$ NS as a function of maximum mass. The dashed black line represents a conservative bound while the red and blue dashed lines represent the actual limit of the dataset.}
    \label{fig:gz_Cs14minmax}
\end{figure}

We use data from about two million piecewise polytropic EOS satisfying causality along with some other important constraints mentioned in \citep{Godzieba:2020bbz} to obtain the phenomenological upper and lower limits on compactness of non-rotating NS. We find that for any given maximum mass there exist upper and lower limit on $C_{max}$, $C^*_{1.6}$ and $C^*_{1.4}$ as shown in Figure \ref{fig:gz_Cminmax}, \ref{fig:gz_Cs16minmax} and \ref{fig:gz_Cs14minmax}. We construct bins in the array of maximum masses and select the minimum and maximum values of radii corresponding to maximum mass as well of 1.4 $M_\odot$ and 1.6 $M_\odot$ NS. 


We use a conservative limit from the data set by constructing a smooothened spline fit with few points. We transform the bounds on radii to the bounds on threshold masses depending upon maximum mass of NS using Eq.~\eqref{eqn:Mth_Rmax_mmax}  plotted in Fig. \ref{fig:kth_rmax_mth}.




\end{document}